\documentclass[12pt]{iopart}

\usepackage{graphicx,color}

\begin{document}

\title[]{Modified saddle-point integral near singularity for the large deviation function  }

\author{Jae Sung Lee$^1$, Chulan Kwon$^2$, and Hyunggyu Park$^1$}

\address{$^1$ School of Physics, Korea Institute for Advanced Study, Seoul 130-722,  Korea}
\address{$^2$ Department of Physics, Myongji University, Yongin, Gyeogii-Do 449-728,  Korea}

\eads{\mailto{jslee@kias.re.kr}, \mailto{ckwon@mju.ac.kr}, \mailto{hgpark@kias.re.kr}}

\begin{abstract}
Long-time-integrated quantities in stochastic processes, in or out of equilibrium, usually exhibit rare but huge fluctuations. Work or heat production is such a  quantity, of which the probability distribution function  displays
an exponential decay characterized by the large deviation function (LDF). The LDF is often deduced from the cumulant generating function through
the inverse Fourier transformation. The saddle-point integration method is a powerful technique to obtain the asymptotic
results in the Fourier integral, but a special care should be taken when the saddle point is located near a singularity of the integrand. In this paper, we present a modified saddle-point method to handle such a difficulty efficiently. We investigate the dissipated and injected heat production in equilibration processes with various initial conditions, as an example, where
the generating functions contain branch-cut singularities as well as power-law ones. Exploiting the new modified saddle-point
integrations, we obtain the leading finite-time corrections for the LDF's, which are confirmed by  numerical results.

\end{abstract}


\maketitle

\section{Introduction}
Detailed balance is satisfied in equilibrium and gives rise to the Boltzmann distribution, which is a well established basis
for equilibrium statistical mechanics. On the other hand, nonequilibrium is characterized by the breakage of detailed balance
and in turn there appears  irreversibility in dynamics. A typical consequence is the existence of nonzero current in state
space. It has been noticed that nonzero current accompanies an incessant production of work, hence heat and
entropy~\cite{Crooks}-\cite{jdnoh1}, each of which satisfies the fluctuation theorem (FT) given at specific initial
distributions~\cite{Evans1}-\cite{Chulan}.
Such time-integrated quantities exhibit rare but huge fluctuations which are  prominent in small systems.
The large deviation function (LDF) is the characteristic function that contains all the information regarding complicated
fluctuations in the long-time limit and has been nowadays one of main issues in nonequilibrium statistical
mechanics~\cite{Zon}-\cite{Tome}.

For a time-integrated quantity $C$ produced from time $t=0$ to $t=\tau$, the LDF $h(\varepsilon)$ for its average production rate $\varepsilon=C/\tau$ is defined as
\begin{equation}
h(\varepsilon) =\lim_{\tau \rightarrow \infty} \frac{1}{\tau} \ln P (\varepsilon;\tau), \label{LDF_definition}
\end{equation}
where $P(\varepsilon;\tau)$ is the probability density function (PDF) of rate $\varepsilon$ for $C$ produced up to time $\tau$. It provides an essential information on the asymptotic property of fluctuations in the long-time limit~\cite{Gallavotti,Lebowitz,Zon,Derrida,Farago,Puglish,Harris,Visco}.

Experimental or numerical confirmation for a theoretically obtained LDF is a very difficult task because the LDF tail
is determined by extremely rare events. Van Zon and Cohen~\cite{Zon} studied heat production of a Brownian particle trapped in the harmonic potential moving with a constant velocity and found that the heat production PDF exhibits a deviation from the conventional FT in the tail region. Their numerical simulation data, however, did not seem to show  good
 accordance with the theoretical LDF  due to an insufficient number of samples. There were also experimental attempts in the electric circuit and mechanical pendulum setups~\cite{Ciliberto}. However, it also seemed not clear that the experimental data are fully
consistent with the theoretical estimates in the tail region.  Therefore, it is desirable to calculate the finite-time correction of the LDF so as to confirm the validity of the theory from the finite-time data in numerical or experimental tests.

The cumulant generating function associated with $P(\varepsilon;\tau)$ is defined as
\begin{equation}
G (\lambda;\tau) = \langle e^{-\lambda\tau \varepsilon} \rangle_\tau = \int d\varepsilon~ P (\varepsilon;\tau) e^{-\lambda\tau \varepsilon}. \label{Fourier_transform}
\end{equation}
In most cases~\cite{Zon,Farago,Visco,Sabhapandit}, it is easier to calculate the generating function than the PDF directly. Then $P(\varepsilon;\tau)$ can be deduced by the inverse Fourier transform of the generating function. The corresponding Fourier integral can be estimated for large $\tau$ as
\begin{equation}
P(\varepsilon;\tau) =\frac{\tau}{2\pi i}\int_{-i\infty}^{i\infty}
d\lambda~ G(\lambda;\tau)e^{\lambda\tau \varepsilon}\simeq \int_{C} d\lambda~ \phi(\lambda) e^{\tau H(\lambda;\varepsilon)} \label{inverse_Fourier}
\end{equation}
where $G(\lambda;\tau)$ is factorized into the exponential term contributed to $H(\lambda;\varepsilon)$ and the leftover to $\phi(\lambda)$ for large $\tau$.
The integral path $C$ is chosen as the steepest descent contour passing through the saddle point, which is usually taken as the solution of $H'(\lambda;\varepsilon)=0$ with $H'=dH/d\lambda$.
We call this saddle point as a {\em conventional} saddle point, denoted by $\lambda_0^* (\varepsilon)$. The Gaussian integration for equation (\ref{inverse_Fourier}) near  $\lambda_0^* (\varepsilon)$, which will be called the conventional saddle-point method, yields
\begin{eqnarray}
P(\varepsilon;\tau) \simeq \sqrt{\frac{2\pi}{\tau |H''(\lambda_0^*;\varepsilon)|}}
~\phi (\lambda_0^*) e^{i\delta} e^{\tau H(\lambda_0^*;\varepsilon)},\label{conventionalSP}
\end{eqnarray}
where $H''=d^2 H/ d\lambda^2$ and $\delta$ is an angle between the steepest descent path and the horizontal axis at $\widetilde{\lambda}_0^*$.
When there is no singularity in the prefactor $\phi(\lambda)$, the above result leads to the correct LDF $h(\varepsilon)$ in equation (\ref{LDF_definition}) and its finite-time correction~\cite{note1}. However, there are many examples where the prefactor
has a power-law type singularity~\cite{Zon,engel,nickelsen,kwon1,kwon2,Hugo}.
If the prefactor has a singularity at $\lambda=\lambda_B$ and $\lambda_0^*(\varepsilon)$ passes through $\lambda_B$ at $\varepsilon=\varepsilon_B$ as $\varepsilon$ varies, the conventional saddle-point method gives rise to the $\delta$-function type divergence in the PDF at $\varepsilon=\varepsilon_B$, which is physically unreasonable. This problem was carefully treated in \cite{Sabhapandit} based on \cite{Wong} and \cite{Sabhapandit1} when the prefactor has a simple or square-root pole, respectively.
However, in those works, the conventional saddle points were still used to construct the steepest descent contour, thus the calculation of the LDF near $\varepsilon=\varepsilon_B$
demands rather complicated algebra as well as composite deformed contours.

In this study, we take a different saddle point, denoted by $\lambda^*(\varepsilon)$, which is the solution of $d[H(\lambda;\varepsilon)+\tau^{-1}\ln \phi(\lambda)]/d\lambda=0$ \cite{Zon}. This saddle point is $\tau$-dependent, but never passes through the singularity. It approaches the singularity only asymptotically in the long-time limit. This feature  simplifies the analysis to obtain the LDF as well as its finite-time correction. However, special care should be taken to calculate the integral near the \emph{modified} saddle point $\lambda^* (\varepsilon)$. When $\lambda^* (\varepsilon)$ approaches the singular point asymptotically, the integration along the steepest descent path near the modified saddle point becomes a {\em non-Gaussian} integral, which means that the usual Gaussian integration cannot be performed as in the conventional saddle point method. This feature was not properly treated
in~\cite{Zon}, where the usual Gaussian integration was used for the modified saddle point. Therefore, we develop a saddle-point integration method to treat a non-Gaussian  integration near this {\em modified} saddle point $\lambda^*(\varepsilon)$, especially when $\lambda^*(\varepsilon)$ asymptotically approaches the singular point.
To illustrate our method explicitly, we revisited the equilibration process~\cite{Lee} as an example. In this case, the prefactor $\phi(\lambda)$ has a square-root singularity with a branch cut. However, our modified method is applicable to general power-law type singularity (see equation~(\ref{alpha-singularity})) and it would be straightforward to generalize to any type of
singularities including an essential singularity. We show all
mathematical details of our modified saddle-point method to obtain the LDF's and their leading  finite-time corrections.

This paper is organized as follows. In Sec.~2, we introduce the modified saddle point method and discuss its advantage
over the conventional method. In Sec.~3, the equilibration process is introduced in brief. In Sec.~4, we calculate the LDF's for the dissipated and injected powers of heat in the long-time limit. In Sec.~5, detailed calculation results are presented for finite-time corrections of the LDF's. We also perform numerical simulations to confirm our results. Numerical data are
in excellent agreement with the analytic results. Finally, we summarize our work in Sec.~6. In the Appendix A, the details
of the modified saddle-point method are presented.

\section{Modified saddle point due to a singularity}\label{sec2}

Consider the case when the prefactor has a singularity such that
\begin{equation}
\phi(\lambda)=\frac{g(\lambda)}{(\lambda-\lambda_B)^{\alpha}} \quad {\rm with} \quad \alpha > 0 \label{alpha-singularity}
\end{equation}
Then, equation (\ref{inverse_Fourier}) becomes
\begin{equation}
P(\varepsilon;\tau)\simeq \int_{C} d\lambda~ \frac{g(\lambda)}{(\lambda-\lambda_B)^\alpha}~ e^{\tau H(\lambda;\varepsilon)}=\int_{C}  d\lambda~ e^{\tau H(\lambda;\varepsilon) -\alpha \ln (\lambda-\lambda_B)+\ln g(\lambda)}, \label{new_saddle_integration}
\end{equation}
where $g(\lambda)$ and $H(\lambda;\varepsilon)$ are analytic functions of $\lambda$. Suppose that the conventional saddle point $\lambda_0^*(\varepsilon)$ satisfying $H'(\lambda;\varepsilon) =0$, passes through $\lambda_B$ at $\varepsilon=\varepsilon_B$,
i.e.~$\lambda_0^*(\varepsilon_B)=\lambda_B$. The modified saddle point $\lambda^*(\varepsilon)$ is determined by the equation, $\tau H'(\lambda;\varepsilon) -\alpha/(\lambda-\lambda_B )+  g'(\lambda)/g(\lambda)=0$.
Here, the last term is always negligible for large $\tau$, thus can be ignored. However, the second term can become comparable to the first term when the saddle point is in the vicinity of $\lambda_B$, and thus, should be taken into account. Hence the saddle point equation becomes
\begin{equation}
S(\lambda;\varepsilon)\equiv H'(\lambda;\varepsilon) -\frac{\alpha}{\tau}\frac{1}{\lambda-\lambda_B}=0. \label{modified_S}
\end{equation}

\begin{figure}[t]
\includegraphics[width=1.0\linewidth]{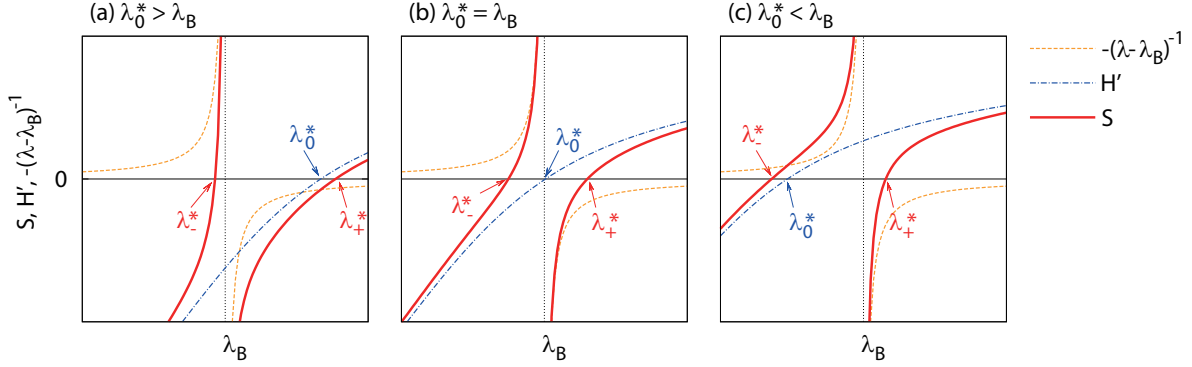}
\caption{(Color online) Locations of $\lambda_0^*$, $\lambda_-^*$, and $\lambda_+^*$ when (a) $\lambda_0^* > \lambda_B$, (b) $\lambda_0^*=\lambda_B$, and (c) $\lambda_0^* < \lambda_B$.
The horizontal axis represents $\lambda$, and the various lines represent
$S$ (thick line), $H'$ (dot-dashed line), and $-(\lambda-\lambda_B)^{-1}$ (dashed line),
respectively.
}\label{modified_saddle_point}
\end{figure}

In contrast to $\lambda_0^*(\varepsilon)$, the solution $\lambda^*(\varepsilon)$ does not pass through $\lambda_B$ but asymptotically approaches it for large $\tau$. In order to illustrate this feature clearly in an example, we assume that $\lambda_B$ and $\lambda_0^*(\varepsilon)$ are located on a real axis and $H'(\lambda;\varepsilon)$ is a real and monotonically increasing function of $\lambda$ on a real axis.
Figs.~\ref{modified_saddle_point}(a), \ref{modified_saddle_point}(b), and \ref{modified_saddle_point}(c) show the plots for $S$, $H'$, and $-\alpha/[\tau(\lambda-\lambda_B)]$ versus real $\lambda$ for $\lambda_0^*>\lambda_B$, $\lambda_0^*=\lambda_B$, and $\lambda_0^*<\lambda_B$ cases, respectively. As shown in the figures, there are always two solutions satisfying $S(\lambda;\varepsilon)=0$; $\lambda_-^*$ and $\lambda_+^*$ which are located on the left and right side of $\lambda_B$ respectively, due to the singularity even though there is only one solution $\lambda_0^*$ for $H\rq{}(\lambda;\varepsilon)=0$. Furthermore, $\lambda_-^*$ and $\lambda_+^*$ cannot pass through $\lambda_B$ as $\varepsilon$ (or $\lambda_0^*$) varies, while $\lambda_0^*(\varepsilon)$ can. Instead, $\lambda_-^*$ and $\lambda_+^*$ asymptotically approach $\lambda_B$ or $\lambda_0^*(\varepsilon)$ for large $\tau$ depending on the location of $\lambda_0^*(\varepsilon)$:
\begin{equation}
\left\{\begin{array}{lll}
\lambda_-^* \rightarrow \lambda_B &\textrm{ and } \quad\lambda_+^* \rightarrow \lambda_0^* (\varepsilon) \quad &\textrm{ when }\lambda_0^*(\varepsilon)>\lambda_B, \nonumber \\
\lambda_-^* \rightarrow \lambda_0^*(\varepsilon)\quad &\textrm{ and } \quad \lambda_+^* \rightarrow \lambda_B &\textrm{ when }  \lambda_0^*(\varepsilon)\leq\lambda_B.
\end{array} \right. \label{two_solutions}
\end{equation}

These modified saddle points, $\lambda_-^*$ and $\lambda_+^*$, make the integration of equation (\ref{new_saddle_integration}) much simpler compared to the case using a conventional one
$\lambda_0^*$. For  general non-integer $\alpha>0$, non-analytic branch cuts appear in the complex plane of $\lambda$, which becomes a nuisance because the integration path should be chosen not to cross them. With the two modified saddle points, it is always possible to choose one of them, of which the steepest descent path does not cross the branch
cuts. In contrast, this is not always possible with a single conventional saddle point.
This is a big advantage of our modified saddle point method over the conventional one.

\begin{figure}[t]
\includegraphics[width=0.6\linewidth]{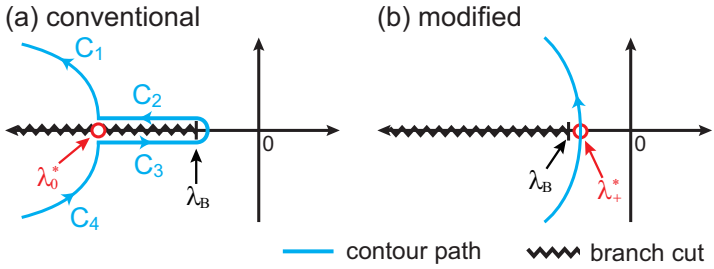}
\caption{(Color online) (a) Contour paths for a conventional saddle point $\lambda_0^*$ when $\lambda_0^*$ is located on a branch cut. (b) Contour path for a modified saddle point $\lambda_+^*$.
}\label{advantage}
\end{figure}

Consider the case of $\lambda_0^*<\lambda_B$ in Fig.~\ref{modified_saddle_point}(c)
with a branch cut on a real axis for $\lambda<\lambda_B$ as in Fig.~\ref{advantage}.
With the choice of $\lambda_+^*$ as the saddle point, one can construct the steepest
descent path not crossing the branch cut in Fig.~\ref{advantage}(b). Hence, the leading contribution of
equation (\ref{new_saddle_integration}) is simply obtained as the integrand evaluated at
$\lambda= \lambda_+^*$:
$P(\varepsilon;\tau)\simeq \exp \left[ \tau H(\lambda_+^*;\varepsilon)- \alpha \ln(\lambda_+^*-\lambda_B)+\ln g(\lambda_+^*) \right]\simeq \exp \left[ \tau H(\lambda_+^*;\varepsilon)\right]$. Note that
$\lambda_+^*-\lambda_B $ vanishes not exponentially, but only algebraically in time $\tau$ (see Appendix A).
From equation (\ref{two_solutions}), we find that the LDF becomes in the infinite-time limit
\begin{equation}
h(\varepsilon)=\left\{\begin{array}{ll}
H(\lambda_0^*(\varepsilon);\varepsilon), & \textrm{ for }  \lambda_0^*(\varepsilon) > \lambda_B  \\
H(\lambda_B;\varepsilon). &\textrm{ for } \lambda_0^*(\varepsilon) \leq \lambda_B ~.
\end{array} \right.\label{LDF0}
\end{equation}
Here, the interesting feature is the `saddle-point fixation' at the singular point $\lambda_B$ for $\lambda_0^*(\varepsilon) \leq \lambda_B$, which is a natural and straightforward consequence in our scheme. Equation (\ref{LDF0}) can be also obtained by using the conventional saddle point scheme. However, in the conventional saddle point scheme, the saddle point lies at the branch cut and nontrivial explanation is needed to derive the same result~\cite{Zon, Farago, Sabhapandit, Sabhapandit1}. This advantage also applies to the case of integer $\alpha$.

Our method has a clear advantage in obtaining a finite-time correction for $P(\varepsilon;\tau)$ when $\lambda_0^*<\lambda_B$.
In the conventional approach~\cite{Farago, Sabhapandit1}, one needs to
construct the detoured path composed of  $C_1$, $C_2$, $C_3$, and $C_4$ segments as in Fig.~\ref{advantage}(a), in order to avoid the branch cut.
Therefore, integrations for all four segments are needed to obtain a finite-time correction. However, in our scheme  (Fig.~\ref{advantage}(b)), we need to calculate only one saddle point integration near $\lambda_+^*$, which makes the formulation much simpler (see equation (\ref{app_int_C3})). In the following sections, we investigate the equilibration process as an example to reveal these advantages more explicitly.

\section{Equilibration process of a Brownian particle}\label{model}

Consider a Brownian particle which is initially in equilibrium with a heat bath A at temperature $T_s$. At $t = 0$, the thermal contact is abruptly switched to the heat bath B at temperature $T_b$ and is maintained forever. For $t\geq 0$, the motion of the Brownian particle with unit mass is described by the Langevin equation
\begin{equation}
\dot{v} = -\gamma v + \xi, \label{Langevin}
\end{equation}
where $v$ is the velocity of the particle and $\gamma$ is the dissipative coefficient. Gaussian white noise $\xi(t)$ satisfies $\left< \xi(t) \right> = 0$ and $\left< \xi(t) \xi(t^\prime) \right> = 2D\delta(t-t^\prime)$, where the fluctuation-dissipation relation holds as $T_b = D/\gamma$. Here, we set the Boltzmann constant $k_B=1$ for convenience. When $T_s = T_b$, the system will remain in the same initial equilibrium state. For $T_s \neq T_b$, however, the system goes through a relaxation process toward equilibrium with the heat bath B. Note that the initial velocity distribution $p_\textrm{in}(v)$ of the particle at $t=0$ is given as
\begin{equation}
p_\textrm{in}(v)=
\sqrt{\frac{\beta\gamma}{2D\pi}}\exp\left(-\frac{\beta\gamma v^2}{2D}\right),
\label{initial_dist}
\end{equation}
where $\beta=T_b/T_s$ is the ratio of the two temperatures.

The total heat in the equilibration process is not an extensive quantity in time because it ceases to be produced as the system approaches equilibrium. However, decomposed into two {\em partial} heats such as dissipated one $-Q_d$ and injected one $Q_i$, then each is accumulated incessantly in time. Multiplying $v$ to equation~(\ref{Langevin}) and integrate it over time $t$ from $0$ to $\tau$, we find
\begin{equation}
\Delta E  = -Q_d + Q_i, \label{heat}
\end{equation}
where $\Delta E=\int_0^\tau dt~ \dot{v}v= \frac{1}{2}v(\tau)^2-\frac{1}{2} v(0)^2$, which is the energy difference between the final and initial time, and partial heats are defined as
\begin{equation}
Q_d \equiv \int_0^\tau dt~ \gamma v^2(t) \quad {\rm and} \quad Q_i\equiv \int_0^\tau dt~ \xi v(t).
\label{heats}
\end{equation}

The fluctuation nature of $Q_d$ and $Q_i$ is quantified by their PDF's, $P(Q_d;\tau)$ and $P(Q_i;\tau)$, respectively. $\left< Q_i \right>$ and $\left< Q_d \right>$ increase linearly in time for sufficiently large $\tau$, while their difference $\left< \Delta E \right> $ is proportional to $ T_b - T_s$ that is bounded. Therefore, in the long-time limit, our conventional wisdom may lead us to expect that the PDF's of $Q_d$ and $Q_i$ will lose all initial memory, thus become independent of $\beta$.
However, it has been noticed in various examples~\cite{Farago,Puglish,Harris,Sabhapandit,Lee} that the effect of initial conditions can remain in the tail of the PDF's (rare-event region) even in the infinite-time limit.

This can be understood intuitively in general as follows:  A large amount of heat with respect to its mean can be produced from a relaxation (decay) dynamics of highly energetic particles, which is an exponentially rare event. Highly energetic particles can be generated from a given initial ensemble at the beginning, which are also exponentially rare but become a source for a large amount of heat in the long-time limit by losing most of their energy by decay dynamics. This initial-condition-dependent rare events certainly affect the (exponentially small) tail of the PDF's even in the infinite-time limit. However, there is another source for highly energetic particles generated by the heat bath, which is also exponentially rare. These two rare events compete each other and sometimes a sharp nontrivial threshold for the initial condition ($\beta$) appears with regard to the initial-condition dependence of the PDF tail shape or the LDF~\cite{Lee}.

In the following sections, we will explicitly calculate the PDF's and analyze their LDF's with finite-time corrections.

\section{Large deviation function}\label{LDF}
The PDF of each heat production $Q$ is expected to exhibit a large deviation nature, $P(Q;\tau)\sim \exp[\tau h(Q/\tau)]$ for large $\tau$. Then, it is convenient to express the LDF as a function of heat production rate, i.e., power. Here, the dissipated power is defined as $\varepsilon_d=Q_d/\tau$ and the injected power as $\varepsilon_i=Q_i/\tau$. Note that $\varepsilon_d\ge 0$ as $Q_d$ is always positive by definition in equation (\ref{heats}), while $\varepsilon_i$ can take any value.

\subsection{Dissipated power: $\varepsilon_d$}\label{ss-Dp}
To calculate $P(\varepsilon_d;\tau)$, it is convenient to compute first its generating function
\begin{equation}
G_d (\lambda;\tau)=\langle e^{-\lambda \tau \varepsilon_d} \rangle_\tau = \int_{-\infty}^{\infty} d\varepsilon_d~ P(\varepsilon_d;\tau) e^{-\lambda \tau \varepsilon_d}. \label{generatingf}
\end{equation}
Equation~(\ref{generatingf}) can be exactly calculated by using the standard path integral method \cite{Farago,Wiegel} with the initial distribution as in equation~(\ref{initial_dist}). The result is given by \cite{Lee}
\begin{eqnarray}
G_d (\lambda;\tau)
&=& e^{\gamma \tau/2} \left( \cosh \eta \gamma \tau + \frac{1+\widetilde{\lambda}/\beta}{\eta} \sinh \eta \gamma \tau \right)^{-1/2}, \label{pi_d}
\end{eqnarray}
where $\widetilde{\lambda} = 2D \lambda/\gamma$ and $\eta = \sqrt{1+2\widetilde{\lambda}}$. Then, the PDF of the dissipated power can be obtained from its inverse Fourier transform as
\begin{eqnarray}
P(\widetilde{\varepsilon}_d;\tau) &=& \frac{\gamma \tau}{4\pi i} \int_{-i\infty}^{i\infty} d \widetilde{\lambda}~ G_d(\gamma\widetilde{\lambda}/2D;\tau) \exp\left(\frac{\gamma \tau \widetilde{\varepsilon}_d \widetilde{\lambda}}{2}\right) \nonumber \\
&=& \frac{\gamma \tau}{4\pi i} \int_{-i\infty}^{i\infty} d \widetilde{\lambda}~ \frac{ e^{\frac{\gamma \tau}{2}(\widetilde{\varepsilon}_d \widetilde{\lambda} + 1)}}{ \sqrt{\cosh \eta \gamma \tau + \frac{1+\widetilde{\lambda}/\beta}{\eta} \sinh \eta \gamma \tau} }, \label{P_d_integration}
\end{eqnarray}
where the dimensionless dissipated power is defined as $\widetilde{\varepsilon}_d = \varepsilon_d/D$. The leading contribution of the above integration can be obtained by using the saddle point method in the large-$\tau$ limit. However, care should be taken when there is a singularity characterized by
\begin{equation}
f_d(\widetilde\lambda;\tau) \equiv \cosh \eta \gamma \tau + \frac{1+\widetilde{\lambda}/\beta}{\eta} \sinh \eta \gamma \tau =0. \label{f_d}
\end{equation}
This singular point is in fact a {\em branch} point connected to the branch cut. For convenience, we choose the branch cut lying on the real-$\tilde\lambda$ axis for $f_d <0$.

\begin{figure}[t]
\includegraphics[width=0.8\linewidth]{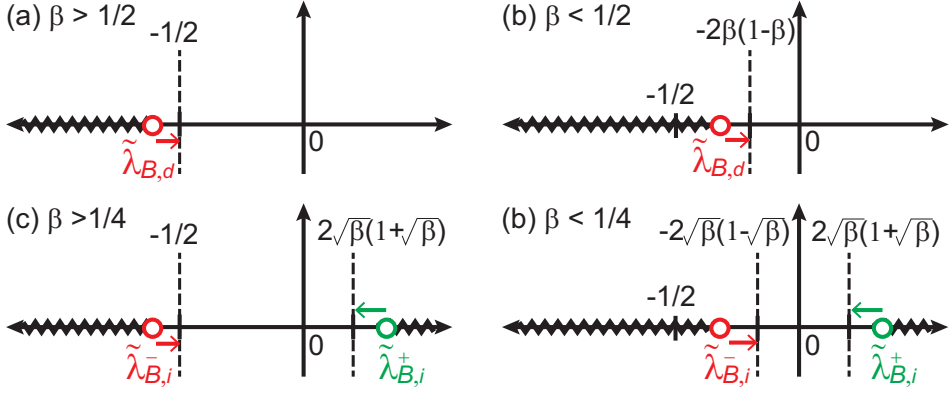}
\caption{(Color online) (a) and (b) show the branch-cut structure of $G_d (\lambda)$ on the complex $\widetilde{\lambda}$ plane for $\beta>1/2$ and $\beta<1/2$, respectively. (c) and (d) show the branch-cut structure of $G_i (\lambda)$ for $\beta>1/4$ and $\beta<1/4$, respectively. Wiggled lines denote branch cuts. A head of an each arrow locates at its asymptotic value of the respective branch point
as $t \rightarrow \infty$.}\label{branch}
\end{figure}

The branch point location depends on $\beta$ in general as shown in figures~\ref{branch}(a) and \ref{branch}(b). First, consider the case with $\beta > 1/2$. For $\widetilde{\lambda}>-1/2$, $\eta$ is positive real, so $f_d \simeq \frac{1}{2} e^{\eta \gamma \tau} (1+ \frac{1+\widetilde{\lambda}/\beta}{\eta})$ for large $\tau$. Here, $1+\widetilde{\lambda}/\beta$ is also positive, so there is no branch point satisfying $f_d=0$.
However, for $\widetilde{\lambda}<-1/2$, $\eta$ becomes pure imaginary and
$f_d=\cos \eta^\prime \gamma \tau + \frac{1+\widetilde{\lambda}/\beta}{\eta^\prime}
\sin \eta^\prime \gamma \tau $ with $\eta^\prime = i\eta$. One can find many solutions
satisfying $f_d=0$ such as $\eta^\prime\gamma\tau\simeq n\pi$ ($n=1,2,\cdots$)
for large $\tau$, equivalently $\tilde\lambda\simeq -\frac{1}{2}[1+(n\pi/\gamma)^2\tau^{-2}]$.
%  $\tilde\lambda\simeq -\frac{1}{2}-n^2\emph{O}(\tau^{-2})$.
So there are infinitely many branching points with vanishingly small intervals between them.
As we choose the branch cut for $f_d<0$, there appear infinitely many patches of branch cuts with vanishingly small sizes in the negative real axis for $\widetilde\lambda<-1/2$.
Note that all branch points have no $\beta$ dependence and approach $\tilde\lambda=-1/2$ from below in the large-$\tau$ limit.
Second, for $\beta<1/2$, $1+\widetilde{\lambda}/\beta$  can be negative for $\widetilde{\lambda}>-1/2$. In this range of $\widetilde{\lambda}$, $\eta$ is positive real and $f_d \simeq \frac{1}{2}e^{\eta \gamma \tau} (1+ \frac{1+\widetilde{\lambda}/\beta}{\eta})$ for large $\tau$. Therefore, the branch point is the solution for $1+ (1+\widetilde{\lambda}/\beta)/\eta=0$, equivalently $\widetilde{\lambda} = -2\beta(1-\beta)$, which is negative and greater than $-1/2$. For $\widetilde{\lambda}<-1/2$, we have a similar
branch-cut structure to the case for $\beta>1/2$.

As discussed in section \ref{sec2}, the new {\em modified} saddle point never passes through the largest branch point (singularity of the prefactor) and its asymptotic location critically depends on the relative position of the conventional saddle point and the largest branch point, see equation (\ref{two_solutions}). Therefore, only the location of the largest branch point is relevant in the saddle point integration. To summarize, the generating function for the dissipated power has infinitely many branch points  with a square-root singularity on the negative real axis and the largest branch point $\widetilde\lambda_{B,d}$ is located asymptotically at
\begin{equation}
\widetilde\lambda_{B,d} =\left\{\begin{array}{ll}
-1/2 &\textrm{ for } \quad \beta>1/2, \nonumber \\
-2\beta(1-\beta) &\textrm{ for } \quad \beta<1/2, \nonumber \\
\end{array} \right. \label{two_branch_points}
\end{equation}
The branch cut is located to the left of the largest branch point on the real axis, so it does not come into play for the saddle point calculation because the modified saddle point always sits outside of the branch cut.

The LDF can be obtained by applying the saddle point approximation to equation~(\ref{P_d_integration}). We  look for a modified saddle point in the region of $\widetilde{\lambda}>-1/2$ (real positive $\eta$). Then, equation~(\ref{f_d}) becomes $f_d\simeq \frac{1}{2}e^{\eta \gamma \tau}\left( 1+\frac{1+\widetilde{\lambda}/\beta}{\eta}\right)$ for large $\tau$. From equation~(\ref{P_d_integration}), the modified saddle point $\widetilde{\lambda}_d^*$ satisfies the following equation:
\begin{equation}
\left.\frac{d}{d\widetilde{\lambda}}\left[\tau H(\widetilde\lambda;\widetilde\varepsilon_d)-\frac{1}{2}\ln \left( 1+\frac{1+\widetilde{\lambda}/\beta}{\eta}\right) \right]\right|_{\widetilde\lambda=\widetilde{\lambda}_d^*} =0, \label{saddle_point_eq}
\end{equation}
where
\begin{equation}
H(\widetilde\lambda;\widetilde\varepsilon_d)=\frac{\gamma }{2}\left( \widetilde{\varepsilon}_d \widetilde{\lambda} +1-\eta \right).
\label{bareLDF_dissipated}
\end{equation}

First, consider the case of $\beta>1/2$. The conventional saddle point $\widetilde{\lambda}_{0,d}^*$ is obtained from $H'(\widetilde\lambda;\widetilde\varepsilon_d)|_{\widetilde\lambda=\widetilde{\lambda}_{0,d}^*}=0$, leading to
\begin{equation}
\widetilde{\lambda}_{0,d}^*(\widetilde{\varepsilon}_d) = -\frac{1}{2}\left( 1-\frac{1}{\widetilde{\varepsilon}_d^2} \right) \label{saddle_point_dissipated_LDF1},
\end{equation}
which is always larger than the branch point $\widetilde{\lambda}_{B,d} = -1/2$ for any $\widetilde{\varepsilon}_d$, similar to Fig.~\ref{modified_saddle_point}(a).
The modified saddle point $\widetilde{\lambda}_{d}^*$ can be obtained from equation (\ref{saddle_point_eq}), which approaches the conventional one $\widetilde{\lambda}_{0,d}^*$ asymptotically
as $\widetilde{\lambda}_d^*(\varepsilon_d)\simeq \widetilde{\lambda}_{0,d}^*(\varepsilon_d) + \emph{O}(\tau^{-1})$.
With no singularity in the vicinity of the saddle point, the conventional saddle-point integration yields the LDF
in the long-time limit (see equation (\ref{LDF0})) for $\beta>1/2$ as
\begin{equation}
h(\widetilde{\varepsilon}_d)=h_1(\widetilde{\varepsilon}_d)\equiv H(\widetilde\lambda_{0,d}^*;\widetilde\varepsilon
_d)= -\frac{\gamma }{4 \widetilde{\varepsilon}_d} (\widetilde{\varepsilon}_d-1)^2. \label{characteristic}
\end{equation}
As $P(\widetilde{\varepsilon}_d)=0$ for $\widetilde{\varepsilon}_d \leq 0$, the LDF is defined only for $\widetilde{\varepsilon}_d > 0$.
This LDF has no initial-temperature or $\beta$ dependence but is determined only by the heat bath properties ($\gamma$ and $D$). Thus, we call equation~(\ref{characteristic}) the heat-bath characteristic curve (HBCC).

For $\beta<1/2$, the branch point is located at $\widetilde{\lambda}_{B,d} = -2\beta(1-\beta) $. The conventional saddle point $\widetilde{\lambda}_{0,d}^*$ is larger than the branch point only when $\widetilde{\varepsilon}_d< (1-2\beta)^{-1}$, and the other way around when $\widetilde{\varepsilon}_d> (1-2\beta)^{-1}$. Therefore, we expect that the LDF is determined by the conventional saddle point for the former case as $h_1(\widetilde{\varepsilon}_d)=H(\widetilde{\lambda}_{0,d}^*;\widetilde\varepsilon_d)$, but by the branch point for the latter case as $h_2(\widetilde{\varepsilon}_d)\equiv H(\widetilde{\lambda}_{B,d};\widetilde\varepsilon_d)$.
To summarize, we find the LDF for $\beta<1/2$ as
\begin{eqnarray}
h(\widetilde{\varepsilon}_d) = \left\{
	\begin{array}{lc}
	h_1(\widetilde{\varepsilon}_d), &{\rm for}\quad \widetilde{\varepsilon}_d < (1-2\beta)^{-1}  \\
	 h_2(\widetilde{\varepsilon}_d)= -\gamma\beta\left[(1-\beta)\widetilde{\varepsilon}_d -1\right] &{\rm for}\quad \widetilde{\varepsilon}_d > (1-2\beta)^{-1} ~.
	\end{array} \right. \label{dissipated_LDF2}
\end{eqnarray}
Note that the LDF is identical to the HBCC for small $\widetilde{\varepsilon}_d$, but the initial condition ($\beta$) dependence shows up in the rare-event region with large $\widetilde{\varepsilon}_d$  when $\beta<1/2$.

As discussed in section \ref{model}, we may expect that the initial-condition dependence remains in the long-time limit when highly energetic particles are prepared in the initial ensemble at high temperature (small $\beta$). Nontheless, it is remarkable to see the sharp and finite threshold ($\beta_c=1/2$) regarding to the existence of the everlasting initial memory in the LDF.
We will investigate the finite-time correction to the LDF in section~\ref{FTC} and Appendix A.

\subsection{Injected power: $\varepsilon_i$}\label{ss-Ip}
In this subsection, we calculate the LDF of the injected power $\varepsilon_i=Q_i/\tau$. The calculation method is similar to that for the dissipated power. The generating function of the injected power is given by \cite{Lee}
\begin{eqnarray}
G_i (\lambda;\tau) = e^{\gamma \tau/2} \left[ \cosh \eta \gamma \tau + \frac{1+\widetilde{\lambda}-\widetilde{\lambda}^2/(2\beta)}{\eta} \sinh \eta \gamma \tau \right]^{-1/2} . \label{pi_i}
\end{eqnarray}
The PDF of the dimensionless injected power $P(\widetilde{\varepsilon}_i)$ with $\widetilde{\varepsilon}_i \equiv \varepsilon_i /D$, can be obtained from the inverse Fourier transform of $G_i(\lambda;\tau)$ as in equation~(\ref{P_d_integration}).

The branch points are determined by the equation
\begin{equation}
f_i(\widetilde{\lambda};\tau)= \cosh \eta \gamma \tau + \frac{1+\widetilde{\lambda} -\widetilde{\lambda}^2/(2\beta)}{\eta} \sinh \eta \gamma \tau =0, \label{f_i}
\end{equation}
which has two relevant solutions: one is on the positive real axis, $\widetilde{\lambda}_{B,i}^+$, and the other is on the negative real axis $\widetilde{\lambda}_{B,i}^-$. For $\widetilde{\lambda}>0$, then $\eta >0$ and $f_i \simeq \frac{1}{2} e^{\eta \gamma \tau} [ 1+ \frac{1+\widetilde{\lambda} -\widetilde{\lambda}^2/(2\beta)}{\eta} ] $ for large $\tau$. It is easy to check that $\widetilde{\lambda}_{B,i}^+ = 2\sqrt{\beta}(1+\sqrt{\beta})$ in the $\tau \rightarrow \infty$ limit and $f_i<0$ for $\widetilde{\lambda}>\widetilde{\lambda}_{B,i}^+$. So it is natural to introduce a branch cut on the real-$\widetilde{\lambda}$ axis to the right of the branch point for $\widetilde{\lambda}>\widetilde{\lambda}_{B,i}^+$.

The negative branch point behaves in a little complicated way as in the case of the dissipated power.
For $\beta>1/4$, $1+\widetilde{\lambda}-\widetilde{\lambda}^2 / (2\beta)$ in equation~(\ref{f_i}) is positive for $-1/2<\widetilde{\lambda}<0$. Then, $f_i$ is always positive and there is no solution satisfying $f_i=0$ within that range
of $\widetilde{\lambda}$. For $\widetilde{\lambda}<-1/2$, $f_i = \cos{\eta' \gamma \tau} + \frac{1+\widetilde{\lambda} -\widetilde{\lambda}^2/(2\beta)}{\eta'} \sin{\eta' \gamma \tau}$ with $\eta'=i\eta$, of which the roots are given by $\eta^\prime\gamma\tau\simeq n\pi$ ($n=1,2,\cdots$) for large $\tau$, equivalently $\widetilde{\lambda}\simeq
-\frac{1}{2}[1+(n\pi/\gamma)^2\tau^{-2}]$, which is independent of $\beta$. For $\beta<1/4$, $1+\widetilde{\lambda}-\widetilde{\lambda}^2 / (2\beta)$ can be negative for $-1/2<\widetilde{\lambda}<0$, then equation~(\ref{f_i}) has a solution in this range, given by $\widetilde{\lambda}_{B,i}^- =-2\sqrt{\beta}(1-\sqrt{\beta})$.

To summarize, the two relevant branch points are located asymptotically at
\begin{equation}
\widetilde\lambda_{B,i}^+=2\sqrt{\beta}(1+\sqrt{\beta}) \quad\quad\quad \textrm{for all}\quad \beta,
\label{branch_point1}
\end{equation}
\begin{equation}
\widetilde\lambda_{B,i}^- =\left\{\begin{array}{ll}
-1/2 &\textrm{ for } \quad \beta>1/4, \nonumber \\
-2\sqrt{\beta}(1-\sqrt{\beta}) &\textrm{ for } \quad \beta<1/4, \nonumber \\
\end{array} \right. \label{branch_point2}
\end{equation}
which are shown in figures~\ref{branch}(c) and \ref{branch}(d).
The branch cut is located for $\widetilde{\lambda}>\widetilde{\lambda}_{B,i}^+$ and $\widetilde{\lambda}<\widetilde{\lambda}_{B,i}^-$ on the real axis.

Now we look for a modified saddle point in between these two branch points. In this region, $\eta$ is always positive and  $f_i \simeq \frac{1}{2} e^{\eta \gamma \tau} [ 1+ \frac{1+\widetilde{\lambda} -\widetilde{\lambda}^2/(2\beta)}{\eta} ] $ for large $\tau$. Then the modified saddle-point $\widetilde{\lambda}_{i}^*$ satisfies the equation
\begin{equation}
\left.\frac{d}{d\widetilde{\lambda}} \left[\tau H(\widetilde{\lambda};\widetilde{\varepsilon}_i)-\frac{1}{2}\ln \left( 1+\frac{1+\widetilde{\lambda}-\widetilde{\lambda}^2/(2\beta)}{\eta}\right) \right]\right|_{\widetilde\lambda=\widetilde{\lambda}_i^*}  =0, \label{saddle_point_eq1}
\end{equation}
where $H(\widetilde\lambda;\widetilde\varepsilon_i)=\frac{\gamma }{2}\left( \widetilde{\varepsilon}_i \widetilde{\lambda} +1-\eta \right)$.

For $\beta>1/4$, the conventional saddle point $\widetilde{\lambda}_{0,i}^*$ is obtained from $H'(\widetilde\lambda;\widetilde\varepsilon_i)|_{\widetilde\lambda=\widetilde{\lambda}_{0,i}^*}=0$, leading to
\begin{equation}
\widetilde{\lambda}_{0,i}^*(\widetilde{\varepsilon}_i) = -\frac{1}{2}\left( 1-\frac{1}{\widetilde{\varepsilon}_i^2} \right) \label{saddle_point_dissipated_LDF1},
\end{equation}
which is always larger than  $\widetilde{\lambda}_{B,i}^- = -1/2$ for any $\widetilde{\varepsilon}_i$, but
may increase and go through $\widetilde{\lambda}_{B,i}^+ = 2\sqrt{\beta}(1+\sqrt{\beta})$ at $\widetilde{\varepsilon}_i= (1+2\sqrt\beta)^{-1}$ as $\widetilde{\varepsilon}_i$ decreases. Therefore, we expect that the LDF is given as
$h_1(\widetilde{\varepsilon}_i)=H(\widetilde{\lambda}_{0,i}^*;\widetilde\varepsilon_i)$ for $\widetilde{\varepsilon}_i> (1+2\sqrt\beta)^{-1}$, and  $h_3(\widetilde{\varepsilon}_i)\equiv H(\widetilde{\lambda}_{B,i}^+;\widetilde\varepsilon_i)$ for $\widetilde{\varepsilon}_i< (1+2\sqrt\beta)^{-1}$. So the LDF is for $\beta>1/4$ as
\begin{eqnarray}
h(\widetilde{\varepsilon}_i) = \left\{
	\begin{array}{lc}
	h_3(\widetilde{\varepsilon}_i)= -\gamma\sqrt\beta\left[1-(1+\sqrt\beta)\widetilde{\varepsilon}_i \right], &{\rm for}\quad \widetilde{\varepsilon}_i < (1+2\sqrt\beta)^{-1}  \\
	 h_1(\widetilde{\varepsilon}_i) &{\rm for}\quad \widetilde{\varepsilon}_i > (1+2\sqrt\beta)^{-1} ~.
	\end{array} \right. \label{injected_LDF1}
\end{eqnarray}

For $\beta<1/4$, we can do a similar analysis to get
$h_4(\widetilde{\varepsilon}_i)\equiv H(\widetilde{\lambda}_{B,i}^-;\widetilde\varepsilon_i)$ for $\widetilde{\varepsilon}_i> (1-2\sqrt\beta)^{-1}$, $h_1(\widetilde{\varepsilon}_i)=H(\widetilde{\lambda}_{0,i}^*;\widetilde\varepsilon_i)$ for $(1+2\sqrt\beta)^{-1}<\widetilde{\varepsilon}_i< (1-2\sqrt\beta)^{-1}$, and  $h_3(\widetilde{\varepsilon}_i)= H(\widetilde{\lambda}_{B,i}^+;\widetilde\varepsilon_i)$ for $\widetilde{\varepsilon}_i< (1+2\sqrt\beta)^{-1}$. So the LDF is for $\beta<1/4$ as
\begin{eqnarray}
h(\widetilde{\varepsilon}_i) = \left\{
	\begin{array}{lc}
	h_3(\widetilde{\varepsilon}_i), &\left( \widetilde{\varepsilon}_i < \frac{1}{1+2\sqrt{\beta}} \right) \\
	h_1(\widetilde{\varepsilon}_i), &\left( \frac{1}{1+2\sqrt{\beta}}< \widetilde{\varepsilon}_i <\frac{1}{1-2\sqrt{\beta}}  \right) \\
	 h_4(\widetilde{\varepsilon}_i)=  -\gamma\sqrt\beta\left[(1-\sqrt\beta)\widetilde{\varepsilon}_i -1\right], &\left( \widetilde{\varepsilon}_i > \frac{1}{1-2\sqrt{\beta}} \right)
	\end{array} \right. \label{injected_LDF2}
\end{eqnarray}

The non-analyticity in the negative tail comes from a trivial constraint that the energy loss of a particle is bounded by its initial energy~\cite{Lee}. However, the positive tail exhibits again the sharp and finite threshold but at the different value of $\beta_c=1/4$. In general, the threshold value should vary with the quantity interested.

\begin{figure}[t]
\includegraphics[width=0.9\linewidth]{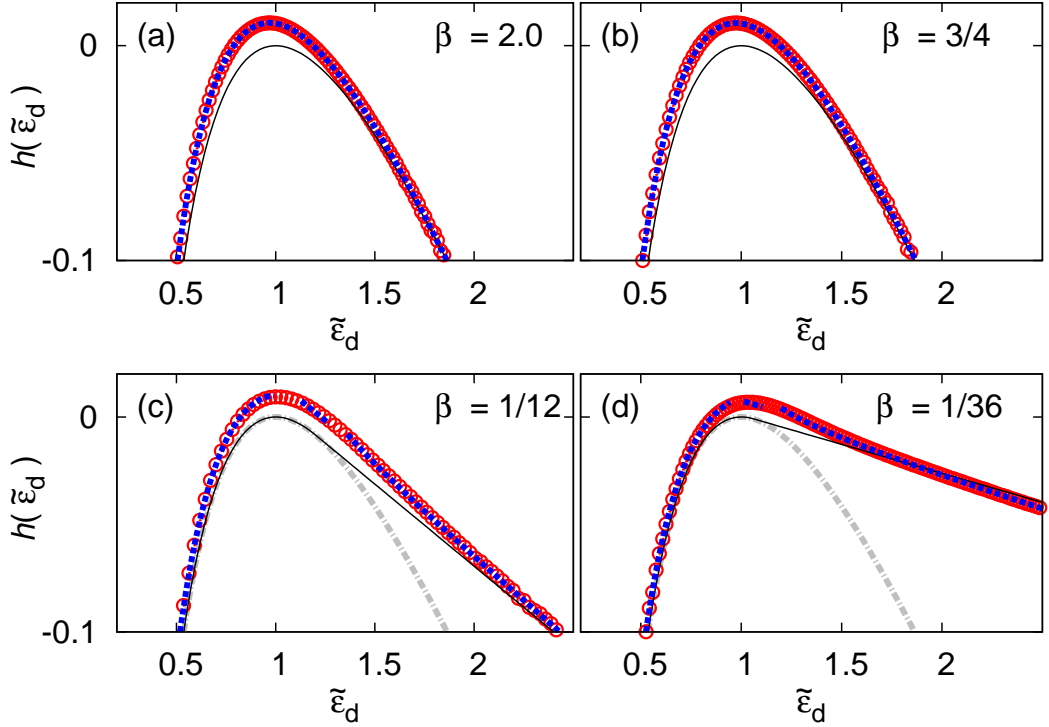}
\caption{(Color online) Panels (a) and (b) show the LDF of the
dissipated power for $\beta = 2.0$ and $\beta =3/4$ ($\beta>1/2$), respectively. The solid line is the HBCC, equation (\ref{characteristic}).
Panels (c) and (d) show the LDF of the dissipated power for $\beta = 1/12$ and $\beta = 1/36$ ($\beta<1/2$), respectively. The solid line denotes equation (\ref{dissipated_LDF2}). In the above four plots, numerical results are presented as open circles in red color and the calculated LDF's with finite-time corrections, given in equations~(\ref{dissipated_LDF1_finite}) and (\ref{dissipated_LDF2_finite}), as thick dotted line in blue color, both obtained at $\tau = 100$. In panels (c) and (d), the dot-dashed line presents the HBCC for comparison.}
\label{finiteLDFs_dissipated}
\end{figure}

\subsection{Comparison with LDF's from numerical simulation results}
We performed numerical simulations for the stochastic differential equation~(\ref{Langevin}) using $\gamma= D=1$ to confirm the LDF's in equations~(\ref{characteristic}), (\ref{dissipated_LDF2}), (\ref{injected_LDF1}), and (\ref{injected_LDF2}) numerically.

Figures~\ref{finiteLDFs_dissipated}(a) and \ref{finiteLDFs_dissipated}(b) show the calculated LDF's of the dissipated power for $\beta >1/2$ cases (black solid line), and the numerically obtained $\tau^{-1}\ln P(\widetilde{\varepsilon}_d)$ (red open circles) at $\tau =100$. As shown in the figures, all the simulation results are close to the HBCC  and independent of $\beta$, as expected from equation~(\ref{characteristic}). Figures~\ref{finiteLDFs_dissipated}(c) and ~\ref{finiteLDFs_dissipated}(d) show the LDF's of the dissipated power for $\beta <1/2$ cases (black solid line) and the numerically obtained $\tau^{-1}\ln P(\widetilde{\varepsilon}_d)$ (red open circles) at $\tau =100$. Contrary to the $\beta>1/2$ cases, the numerical results depends on $\beta$ for $\widetilde{\varepsilon}_d > 1/(1-2\beta)$, as expected from equation~(\ref{dissipated_LDF2}). There appear noticeable deviations of numerical data from the calculated LDF curves due to finite-time effect, which are resolved in section~5.

\begin{figure}[t]
\includegraphics[width=0.9\linewidth]{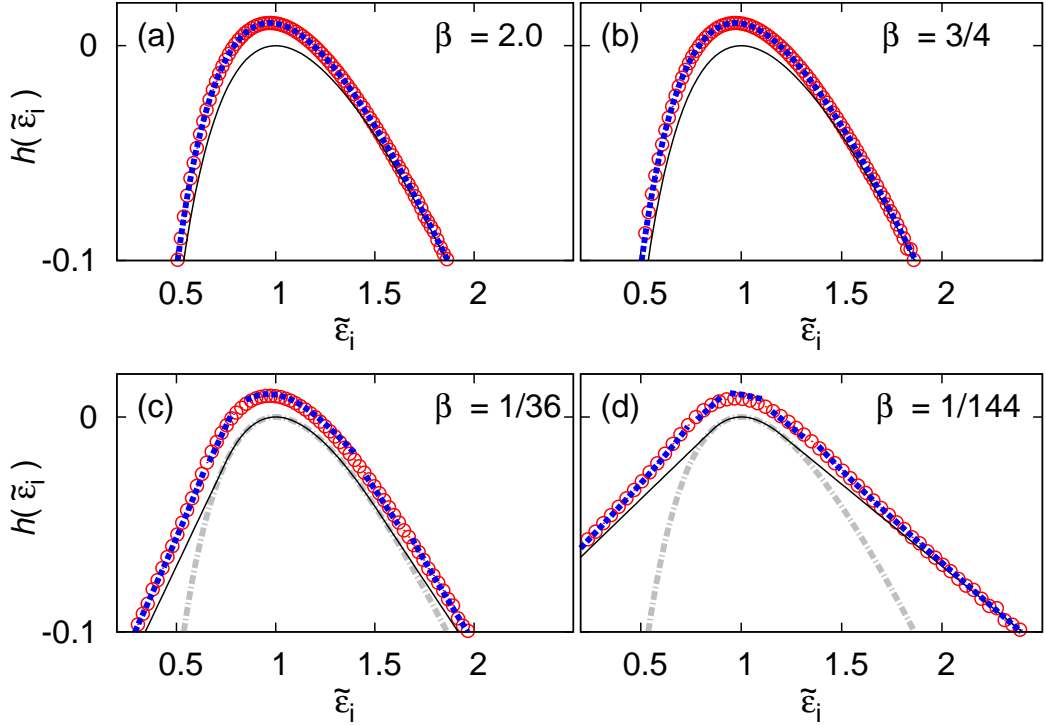}
\caption{(Color online) Panels (a) and (b) show the LDF of the
injected power for $\beta = 2.0$ and $\beta =3/4$ ($\beta>1/4$), respectively. The solid line plots equation~(\ref{injected_LDF1}). Panels (c) and (d) show the LDF of the injected power for $\beta = 1/36$ and $\beta = 1/144$ ($\beta<1/4$), respectively. The solid line plots equation~(\ref{injected_LDF2}). In the above four plots, numerical results are presented as open circles in red color and the calculated LDF's with finite-time corrections at $\tau = 100$, given in equations~ (\ref{injected_LDF1_finite}) and (\ref{injected_LDF2_finite}), as thick dotted line in blue color. In panels (c) and (d), we plot the HBCC as the dot-dashed line for comparison.
}\label{finiteLDFs_injected}
\end{figure}

Figure~\ref{finiteLDFs_injected}(a) and \ref{finiteLDFs_injected}(b) show the analytical (black solid line) and numerical (red open circles) LDF's of the injected power for $\beta>1/4$ at $\tau =100$. In these figures, the $\beta$-dependent LDF for $\widetilde{\varepsilon}_i <(1+2\sqrt{\beta})^{-1}$ does not appear simply because the region is outside of the plot range. Figure~\ref{finiteLDFs_injected}(c) and \ref{finiteLDFs_injected}(d) show the analytical (black solid line) and numerical (red open circles) LDF's of the injected power for $\beta<1/4$. In these cases, the three regions are clearly seen, as expected from equation~(\ref{injected_LDF2}). Similar deviations are seen due to finite-time effects.

\section{Finite-time corrections}\label{FTC}

In this section, we calculate the LDF's for the dissipated and injected powers with finite-time corrections up to the first order
in $\tau^{-1}$ from the PDF integrations given by (see equation~(\ref{P_d_integration}))
\begin{equation}
P(\widetilde{\varepsilon}_{x};\tau) = \frac{\gamma \tau}{4\pi i} \int_{-i\infty}^{i\infty} d
\widetilde{\lambda}~ G_{x}(\gamma\widetilde{\lambda}/2D;\tau) \exp\left(\frac{\gamma \tau \widetilde{\varepsilon}_{x}
\widetilde{\lambda}}{2}\right), \label{PDF_x_integration}
\end{equation}
where $x$ is $d$ (dissipated) or $i$ (injected), and $G_{x}(\gamma\widetilde{\lambda}/2D;\tau)$ is its generating function
of equation (\ref{pi_d}) or (\ref{pi_i}). Factorizing $G_x$ into the exponential term and the leftover as in equation~(\ref{inverse_Fourier}), we  write the above PDF integration as
\begin{equation}
P(\widetilde{\varepsilon}_x;\tau) = \int_{-i\infty}^{i\infty} d\widetilde{\lambda}~ \phi_x (\widetilde{\lambda})~ e^{\tau H(\widetilde{\lambda};\widetilde{\varepsilon}_x)},
\label{P_d_large}
\end{equation}
where $H(\widetilde{\lambda};\widetilde{\varepsilon}_x)=\frac{\gamma}{2}(\widetilde{\varepsilon}_x\widetilde{\lambda}+1-\eta)$
with real $\eta=\sqrt{1+2 \widetilde{\lambda}}$.

We employ the modified saddle-point integration to evaluate the PDF's and their
finite-time corrections in Appendix A, where the prefactor function $\phi_x(\widetilde{\lambda})$ has a power-law
singularity like in equation (\ref{alpha-singularity}). In the following subsections, we present the results for
various cases, especially focussing on the closed form of the LDF's in the scaling regimes.
We use the notations $\widetilde{\lambda}_{B,x}$ (branch point), $\widetilde{\lambda}_{0,x}^*$ (conventional
saddle point), and $\widetilde{\lambda}_{x}^*$ (modified saddle point) with their relative distances
$\Delta\widetilde{\lambda}_{x}=\widetilde{\lambda}_{0,x}^*-\widetilde{\lambda}_{B,x}$ and
$\delta\widetilde{\lambda}_{B,x}=\widetilde{\lambda}_{x}^*-\widetilde{\lambda}_{B,x}$.

\subsection{Dissipated power}

The prefactor reads $\phi_d(\widetilde{\lambda})
=\frac{\sqrt{2}\gamma \tau}{4 \pi i}[1+\frac{1+\widetilde{\lambda}/\beta}{\eta}]^{-1/2}
=\frac{\sqrt{2}\gamma \tau}{4 \pi i}\left[\frac{ 2\beta\eta}{\left(\eta+2\beta-1 \right)(\eta+1)}\right]^{1/2}$
with $\eta=\sqrt{1+2 \widetilde{\lambda}}$.

\subsubsection{$\beta >1/2$ case :}

In subsection \ref{ss-Dp}, we found $\widetilde{\lambda}_{B,d}=-\frac{1}{2}$ and $\widetilde{\lambda}_{0,d}^*(\widetilde{\varepsilon}_d)=-\frac{1}{2}(1-1/\widetilde{\varepsilon}_d^2)$, leading to
$\Delta\widetilde{\lambda}_{d}>0$ for all $\widetilde{\varepsilon}_d$. So we can apply the conventional saddle point integration result in equation (\ref{app_conventional_finite_int}).
Using $H''(\widetilde{\lambda}_{0,d}^*;\widetilde{\varepsilon}_d)=\frac{\gamma}{2} \widetilde{\varepsilon}_d^3$, $\phi_d(\widetilde{\lambda}_{0,d}^*)=\frac{\gamma \tau}{2\pi i}\left[ \frac{1}{\beta\widetilde{\varepsilon}_d} (\widetilde{\varepsilon}_d+1) \{(2\beta-1)\widetilde{\varepsilon}_d+1 \}  \right]^{-1/2}$, and equation
(\ref{characteristic}) for $H(\widetilde{\lambda}_{0,d}^*;\widetilde{\varepsilon}_d)$, then the LDF  becomes
\begin{equation}
h(\widetilde{\varepsilon}_d) = h_1(\widetilde{\varepsilon}_d) + \frac{1}{\tau} r_{1} (\widetilde{\varepsilon}_d), ~~~(\widetilde{\varepsilon_d}>0)
\label{dissipated_LDF1_finite}
\end{equation}
where
\begin{equation}
r_{1} (\widetilde{\varepsilon}_d) = \frac{1}{2} \ln \left[
\frac{\gamma \tau\beta}{\pi {\widetilde{\varepsilon}_d}^2 (\widetilde{\varepsilon}_d+1) [(2\beta-1)\widetilde{\varepsilon}_d+1]}~\right].
\end{equation}
The above equation with $\tau = 100$ is presented in Figs.~\ref{finiteLDFs_dissipated}(a) and \ref{finiteLDFs_dissipated}(b)
for $\beta = 2.0$ and $3/4$ cases, respectively. They are in excellent accord with the numerical data.

\subsubsection{$\beta <1/2$ case :}

In subsection \ref{ss-Dp}, we found that the branch point location depends on $\beta$, given by $\widetilde{\lambda}_{B,d}=-2\beta(1-\beta)$ with the same $\widetilde{\lambda}_{0,d}^*(\widetilde{\varepsilon}_d)$ as above.
Note that $\Delta\widetilde{\lambda}_{d}>0$ for $\widetilde{\varepsilon}_d <(1-2\beta)^{-1}$, while
$\Delta\widetilde{\lambda}_{d}<0$ otherwise. The prefactor is written as
\begin{equation}
\phi_d(\widetilde{\lambda})= \frac{g_d(\widetilde{\lambda})}{\left({\widetilde{\lambda}-\widetilde{\lambda}_{B,d}}\right)^{1/2}}~,
\end{equation}
where the analytic function $g_d(\widetilde{\lambda})$ is given by
\begin{equation}
g_d(\widetilde{\lambda})=\frac{\sqrt{2}\gamma \tau}{4 \pi i} \left[\frac{ \beta\eta\left(\eta+1-2\beta \right)}{ \eta+1}\right]^{1/2} \label{g_lambda1}
\end{equation}
with $\eta=\sqrt{1+2 \widetilde{\lambda}}$.

For $\widetilde{\varepsilon}_d >(1-2\beta)^{-1}$,
we use equation (\ref{app_int_C3}) with
  $g_d(\widetilde{\lambda}_{B,d})=\frac{\sqrt{2}\gamma \tau}{4\pi i} (1-2\beta)\left(\frac{\beta}{1-\beta} \right)^{1/2}$, $H'(\widetilde{\lambda}_{B,d};\widetilde{\varepsilon}_d)=\frac{\gamma}{2} (\widetilde{\varepsilon}_d-\frac{1}{1-2\beta})$,
$H''(\widetilde{\lambda}_{B,d};\widetilde{\varepsilon}_d)=\frac{\gamma}{2} \frac{1}{(1-2\beta)^3}$,
$H(\widetilde{\lambda}_{B,d};\widetilde{\varepsilon}_d)$ in equation~(\ref{dissipated_LDF2}),
and $\delta\widetilde{\lambda}_{B,d}$ in equation (\ref{app_delta_z_3}) to generate the accurate numerical data.
For $\widetilde{\varepsilon}_d <(1-2\beta)^{-1}$, we can use equation (\ref{app_int_C32}) with
$g_d(\widetilde{\lambda}_{0,d}^*)=\frac{\sqrt{2}\gamma \tau}{4\pi i} \left[\frac{\beta ((1-2\beta)\widetilde{\varepsilon}_d+1)}{\widetilde{\varepsilon}_d(\widetilde{\varepsilon}_d+1)} \right]^{1/2}$,
$H''(\widetilde{\lambda}_{0,d}^*;\widetilde{\varepsilon}_d)=\frac{\gamma}{2} \widetilde{\varepsilon}_d^3$,
$H(\widetilde{\lambda}_{0,d}^*;\widetilde{\varepsilon}_d)$ in equation~(\ref{dissipated_LDF2}),
and $\delta\widetilde{\lambda}_{B,d}\approx \Delta\widetilde{\lambda}_{d}$.

The LDF with a finite-time correction in the scaling regimes can be obtained from equation~(\ref{app_I_tot1}) as
\begin{eqnarray}
h(\widetilde{\varepsilon}_d) = \left\{
	\begin{array}{lc}
	h_1(\widetilde{\varepsilon}_d) +\frac{1}{\tau} r_{1} (\widetilde{\varepsilon}_d), &  \frac{1}{1-2\beta}-\widetilde{\varepsilon}_d\gg \tau^{-1/2}   \\
	h_2(\widetilde{\varepsilon}_d) + \frac{1}{\tau} r_2 (\widetilde{\varepsilon}_d), & \left| \widetilde{\varepsilon}_d - \frac{1}{1-2\beta} \right| \ll \tau^{-1/2} \\
	h_2(\widetilde{\varepsilon}_d) +\frac{1}{\tau} r_3 (\widetilde{\varepsilon}_d), &  \widetilde{\varepsilon}_d - \frac{1}{1-2\beta} \gg \tau^{-1/2}
	\end{array} \right. \label{dissipated_LDF2_finite}
\end{eqnarray}
where
\begin{eqnarray}
r_2 (\widetilde{\varepsilon}_d) &=& \frac{1}{2}\ln \left[\frac{(\gamma \tau)^{3/2}(2\beta)(1-2\beta)^{7/2} \Gamma^2 (5/4)}{\pi^2
 (1-\beta)}~\right], \nonumber \\
r_3 (\widetilde{\varepsilon}_d) &=& \frac{1}{2} \ln \left[\frac{\gamma \tau\beta(1-2\beta)^2 }{\pi
 (1-\beta)\left(\widetilde{\varepsilon}_d - 1/(1-2\beta)\right)}~\right]. \nonumber
\end{eqnarray}
The above equation with $\tau = 100$ is compared with the numerical data in Figs.~\ref{finiteLDFs_dissipated}(c) and \ref{finiteLDFs_dissipated}(d) for $\beta = 1/12$ and $1/36$ cases, respectively. As shown in the figures,
they show excellent agreement with numerical data.

\subsection{injected power}

The prefactor reads $\phi_i(\widetilde{\lambda})=\frac{\sqrt{2}\gamma \tau}{4 \pi i}\left[1+\frac{1+\widetilde{\lambda}-
\widetilde{\lambda}^2/(2\beta)}{\eta}\right]^{-1/2}=\frac{\sqrt{2}\gamma \tau}{4 \pi i (\eta+1)}\left[\frac{ 8\beta\eta}{4\beta-(\eta-1)^2 }\right]^{1/2}$.

\subsubsection{$\beta >1/4$ case :}

As seen in subsection \ref{ss-Ip}, two branch points are located at $\widetilde{\lambda}_{B,i}^-=-\frac{1}{2}$ and $\widetilde{\lambda}_{B,i}^+=2\sqrt{\beta}(1+\sqrt{\beta})$, respectively.
As $\widetilde{\lambda}_{0,i}^* = -\frac{1}{2}(1-1/\widetilde{\varepsilon}_i^2)$,
$\Delta \widetilde{\lambda}_{i}^->0$ for all $\widetilde{\varepsilon}_i$, but
$\Delta \widetilde{\lambda}_{i}^+$ changes its sign at $\widetilde{\varepsilon}_i=(1+2\sqrt{\beta})^{-1}$.
Thus, we can write the prefactor $\phi_i(\widetilde{\lambda})$ as
\begin{equation}
\phi_i(\widetilde{\lambda})= \frac{g_{i}^+(\widetilde{\lambda})}{(\widetilde{\lambda}_{B,i}^+ - \widetilde{\lambda})^{{1}/{2}}},
\end{equation}
where the analytic function $g_{i}^+(\widetilde{\lambda})$ is given by
\begin{equation}
g_{i}^+ (\widetilde{\lambda})=\frac{\sqrt{2}\gamma \tau}{4 \pi i (\eta+1)}
\left[ \frac{4\beta\eta (\eta+1+2\sqrt\beta)} {\eta-1+2\sqrt\beta}\right]^{{1}/{2}}.
\end{equation}

For $\widetilde{\varepsilon}_i <(1+2\sqrt\beta)^{-1}$,
we use the slightly modified version of equation (\ref{app_int_C3}) by replacing
$H'$ by $-H'$ and $\delta z_B$ by $-\delta z_B$, because the modified saddle point is located to the left of the
branching point. In order to generate the accurate numerical data, we take
  $g_i^+(\widetilde{\lambda}_{B,i}^+)=\frac{\gamma \tau}{4\pi i} \beta^{1/4}\left[\frac{1+2\sqrt{\beta}}{1+\sqrt{\beta}}\right]$,
  $H'(\widetilde{\lambda}_{B,i}^+;\widetilde{\varepsilon}_i)=\frac{\gamma}{2} \left(\widetilde{\varepsilon}_i-\frac{1}{1+2\sqrt{\beta}} \right)$,
$H''(\widetilde{\lambda}_{B,i}^+;\widetilde{\varepsilon}_i)=
\frac{\gamma}{2} \frac{1}{\left(1+2\sqrt{\beta} \right)^3}$,
$H(\widetilde{\lambda}_{B,i}^+;\widetilde{\varepsilon}_i)$ in equation~(\ref{injected_LDF1}),
and $\delta\widetilde{\lambda}_{B,i}^+$ in equation (\ref{app_delta_z_3}).
For $\widetilde{\varepsilon}_i >(1+2\sqrt\beta)^{-1}$, we also use the slightly modified version of
equation (\ref{app_int_C32}) by replacing $\delta z_B$ and $\Delta z$ by $-\delta z_B$ and $-\Delta z_B$, respectively.
We take $g_{i}^+(\widetilde{\lambda}_{0,i}^*)=\frac{\sqrt{2}\gamma \tau}{4\pi i} \left[\frac{4\beta\widetilde{\varepsilon}_i(1+(2\sqrt{\beta}+1)\widetilde{\varepsilon}_i )}{(1+\widetilde{\varepsilon}_i)^2
(1 +(2\sqrt{\beta}-1)\widetilde{\varepsilon}_i )} \right]^{1/2}$,
$H''(\widetilde{\lambda}_{0,i}^*;\widetilde{\varepsilon}_i)=\frac{\gamma}{2} \widetilde{\varepsilon}_i^3$,
 $H(\widetilde{\lambda}_{0,i}^*;\widetilde{\varepsilon}_i)$ in equation~(\ref{injected_LDF1}),
and  $\delta\widetilde{\lambda}_{B,i}^+\approx \Delta\widetilde{\lambda}_{i}^+$.

The LDF with a finite-time correction can be obtained from equation~(\ref{app_I_tot1})
with $H'\rightarrow -H'$ and $\Delta z\rightarrow -\Delta z$ as
\begin{eqnarray}
h(\widetilde{\varepsilon}_i) = \left\{
	\begin{array}{lc}
	h_3(\widetilde{\varepsilon}_i) +\frac{1}{\tau} r_4 (\widetilde{\varepsilon}_i), &  \frac{1}{1+2\sqrt{\beta}}-\widetilde{\varepsilon}_i  \gg \tau^{-1/2}\\
	h_3(\widetilde{\varepsilon}_i) + \frac{1}{\tau} r_5 (\widetilde{\varepsilon}_i), &  \left| \widetilde{\varepsilon}_i - \frac{1}{1+2\sqrt{\beta}} \right| \ll \tau^{-1/2} \\
	h_1(\widetilde{\varepsilon}_i) +\frac{1}{\tau} r_6 (\widetilde{\varepsilon}_i), &  \widetilde{\varepsilon}_i - \frac{1}{1+2\sqrt{\beta}} \gg \tau^{-1/2}
	\end{array} \right.  \label{injected_LDF1_finite}
\end{eqnarray}
where
\begin{eqnarray}
r_4 (\widetilde{\varepsilon}_i) &=& \frac{1}{2} \ln\left[
\frac{\gamma \tau\sqrt{\beta} (1+2\sqrt{\beta})^2 }{2\pi (1+\sqrt{\beta})^2\left(1/(1+2\sqrt{\beta})-\widetilde{\varepsilon}_i \right)}~ \right], \nonumber \\
r_5 (\widetilde{\varepsilon}_i) &=& \frac{1}{2}\ln \left[\frac{(\gamma \tau)^{3/2}\sqrt\beta(1+2\sqrt\beta)^{7/2} \Gamma^2 (5/4)}{\pi^2
 (1+\sqrt\beta)^2}~\right], \nonumber \\
r_6 (\widetilde{\varepsilon}_i) &=& \frac{1}{2}\ln\left[
\frac{\gamma \tau (4 \beta)}{\pi (\widetilde{\varepsilon}_i+1)^2 [(4\beta-1){\widetilde{\varepsilon}_i}^2+2\widetilde{\varepsilon}_i-1]}~\right], \nonumber
\end{eqnarray}
The above equation with $\tau = 100$ is compared with numerical data in Figs.~\ref{finiteLDFs_injected}(a) and \ref{finiteLDFs_injected}(b) for $\beta = 2.0$ and $\beta= 3/4$, respectively.

\subsubsection{$\beta <1/4$ case :}

As seen in subsection \ref{ss-Ip}, two branch points are located at $\widetilde{\lambda}_{B,i}^-=-2\sqrt{\beta}(1-\sqrt{\beta})$ and $\widetilde{\lambda}_{B,i}^+=2\sqrt{\beta}(1+\sqrt{\beta})$, respectively. Note that $\widetilde{\lambda}_{B,i}^-$ is now
larger than $-\frac{1}{2}$ and the conventional saddle point $\widetilde{\lambda}_{0,i}^*$ passes over to the left side of the branch point $\widetilde{\lambda}_{B,i}^-$  for $\Delta \widetilde{\lambda}_{i}^- <0$. Thus, the LDF should be the same as in equation (\ref{injected_LDF1_finite}) for $\Delta \widetilde{\lambda}_{i}^- >0$ or equivalently
$\widetilde{\varepsilon}_i < (1-2\sqrt\beta)^{-1}$, but one needs to focus on the singularity behavior of the prefactor near
$\widetilde{\lambda}\sim \widetilde{\lambda}_{B,i}^-$ to analyze the LDF for
$\widetilde{\varepsilon}_i > (1-2\sqrt\beta)^{-1}$.
We can rewrite the prefactor $\phi_i(\widetilde{\lambda})$ as
\begin{equation}
\phi_i(\widetilde{\lambda})= \frac{g_{i}^-(\widetilde{\lambda})}{(\widetilde{\lambda} - \widetilde{\lambda}_{B,i}^-)^{{1}/{2}}},
\end{equation}
where the analytic function $g_{i}^-(\widetilde{\lambda})$ is given by
\begin{equation}
g_{i}^- (\widetilde{\lambda})=\frac{\sqrt{2}\gamma \tau}{4 \pi i (\eta+1)}
\left[ \frac{4\beta\eta (\eta+1-2\sqrt\beta)} {1+2\sqrt\beta -\eta}\right]^{{1}/{2}}.
\end{equation}

For $\widetilde{\varepsilon}_i >(1-2\sqrt\beta)^{-1}$,
we use  equation (\ref{app_int_C3}) with
  $g_i^-(\widetilde{\lambda}_{B,i}^-)=\frac{\gamma \tau}{4\pi i} \beta^{1/4}\left[\frac{1-2\sqrt{\beta}}{1-\sqrt{\beta}}\right]$,
  $H'(\widetilde{\lambda}_{B,i}^-;\widetilde{\varepsilon}_i)=\frac{\gamma}{2} \left(\widetilde{\varepsilon}_i-\frac{1}{1-2\sqrt{\beta}} \right)$,
$H''(\widetilde{\lambda}_{B,i}^-;\widetilde{\varepsilon}_i)=
\frac{\gamma}{2} \frac{1}{\left(1-2\sqrt{\beta} \right)^3}$,
$H(\widetilde{\lambda}_{B,i}^-;\widetilde{\varepsilon}_i)$ in equation~(\ref{injected_LDF2}),
and $\delta\widetilde{\lambda}_{B,i}^-$ in equation (\ref{app_delta_z_3}).
Then, the LDF with a finite-time correction becomes
\begin{eqnarray}
h(\widetilde{\varepsilon}_i) = \left\{
	\begin{array}{ll}
	h_3(\widetilde{\varepsilon}_i) +\frac{1}{\tau} r_4 (\widetilde{\varepsilon}_i), & \frac{1}{1+2\sqrt{\beta}}-\widetilde{\varepsilon}_i  \gg \tau^{-1/2} \\
	h_3(\widetilde{\varepsilon}_i) + \frac{1}{\tau} r_5 (\widetilde{\varepsilon}_i), &  \left| \widetilde{\varepsilon}_i - \frac{1}{1+2\sqrt{\beta}} \right| \ll \tau^{-1/2} \\
	h_1(\widetilde{\varepsilon}_i) +\frac{1}{\tau} r_6 (\widetilde{\varepsilon}_i), &  \widetilde{\varepsilon}_i-\frac{1}{1+2\sqrt{\beta}} \gg \tau^{-1/2}~~\textrm{and}~~ \frac{1}{1-2\sqrt{\beta}}-\widetilde{\varepsilon}_i
\gg \tau^{-1/2}\\
	h_4(\widetilde{\varepsilon}_i) +\frac{1}{\tau} r_7 (\widetilde{\varepsilon}_i), &  \left| \widetilde{\varepsilon}_i - \frac{1}{1-2\sqrt{\beta}} \right| \ll \tau^{-1/2} \\
	h_4(\widetilde{\varepsilon}_i) + \frac{1}{\tau} r_8 (\widetilde{\varepsilon}_i), & 	 \widetilde{\varepsilon}_i - \frac{1}{1-2\sqrt{\beta}} \gg \tau^{-1/2}
	\end{array} \right.  \label{injected_LDF2_finite}
\end{eqnarray}
where
\begin{eqnarray}
r_7 (\widetilde{\varepsilon}_i) &=& \frac{1}{2}\ln \left[\frac{(\gamma \tau)^{3/2}\sqrt\beta(1-2\sqrt\beta)^{7/2} \Gamma^2 (5/4)}{\pi^2  (1-\sqrt\beta)^2}~\right], \nonumber \\
r_8 (\widetilde{\varepsilon}_i) &=& \frac{1}{2} \ln\left[
\frac{\gamma \tau\sqrt{\beta} (1-2\sqrt{\beta})^2 }{2\pi (1-\sqrt{\beta})^2\left(\widetilde{\varepsilon}_i - 1/(1-2\sqrt{\beta})\right)}~ \right] . \nonumber
\end{eqnarray}
In Figs.~\ref{finiteLDFs_injected}(c) and \ref{finiteLDFs_injected}(d), the above equation with $\tau = 100$  is compared
with numerical data for $\beta = 1/36$ and $\beta= 1/144$, respectively.

\section{Summary}

In this study, we introduced a new method to manipulate a saddle-point integral when a saddle point is in the vicinity of a singular point characterized by a power-law singularity as $(\lambda-\lambda_B)^{-\alpha}$ with $\alpha>0$. Instead of the integration around the conventional saddle point, we choose a modified saddle point, which takes into account the singularity of the prefactor in the long-time limit. The main feature of the modified saddle point is that it does not pass through the singularity as the parameter varies, but asymptotically approaches it for the long-time limit, while the conventional
saddle point passes through it, leading to a rather complicated integration. This property results in the so-called `saddle-point fixation' which simplifies the analysis to obtain a leading-order term as well as finite-time corrections of the integral.

To obtain leading finite-time corrections, one should do the integral near the modified saddle point, which turns out to be
 non-Gaussian in general, especially when the modified saddle point is asymptotically close to the singular point. However,
 it can be still explicitly written in a specific non-Gaussian integral form, which can be evaluated numerically with very high precision. In general, there exist three different scaling regimes depending on the relative position of the conventional saddle point and the singularity, where the non-Gaussian integral can be done analytically.

To explicitly show the mathematical convenience of our method, we investigated the PDF's of the dissipated and injected heats for a Brownian particle whose initial state is not in equilibrium with a thermal bath.
The PDF of the power $\widetilde{\varepsilon}$ of each heat can be found from the inverse Fourier transform of the generating function. The corresponding generating function has a square-root singularity with $\alpha=1/2$.
By using our method, we obtained the LDF's and its finite-time corrections for each heat. We found that there are sharp transitions of the LDF's depending on the ratio of the initial and heat-bath temperatures. From the numerical simulations, we also confirmed that our method gives correct finite-time corrections.

The mathematical finding in this paper is not specific to the square-root singularity, but applies to general $\alpha$.
Recently, we have investigated the problem where the harmonic potential is pulled by a constant speed and the particle is prepared with the initial distribution at the temperature different from the heat bath temperature. The motion in $d$ space dimensions gives rise to the pole of order $d/2$ in the generating function. We have observed the similar asymptotic behavior of the saddle-point and found the PDF, hence the LDF, by using the modified saddle-point integral technique developed here~\cite{kwangmoo}.

The PDF obtained from the experimental and simulation data accumulated for a long time contains strong statistical errors, especially in the tail of rare events. Therefore, it is a desirable task to find the LDF with finite-time correction analytically. The LDF with finite-time corrections obtained from the modified saddle-point integral technique shows an excellent agreement with the simulation data. We expect that our method would be useful for many similar problems with a singular generating function, which appears frequently in various non-equilibrium phenomena \cite{Zon, Sabhapandit, Lee}.

\ack
This research was supported by the NRF Grant No.~2011-
35B-C00014(J.S.L.), 2013R1A1A2011079(C.K.), and 2013R1A1A2A10009722(H.P.).

\appendix
\section{Modified saddle-point integration near singularity}
Consider the following integral:
\begin{equation}
I(\varepsilon) = \int_{-i \infty}^{i \infty} dz \frac{g(z)}{(z-z_B)^\alpha} e^{\tau H(z;\varepsilon)}~, \label{app_int}
\end{equation}
where $\tau$ is a large positive number, $g(z)$ is an analytic function of $z$ and, $\alpha$ is a positive number. Note that
$\alpha=1/2$ for the equilibration process we studied in the main text. For convenience, we consider that
the singularity at $z=z_B$ lies on the negative real axis of $z$ and  the branch cut is chosen to its left side on the real axis as shown in Fig.~\ref{integration_schematic}. It would be straightforward to apply our procedure to other general cases. We assume that
$H(z;\varepsilon)$ is an analytic function of $z$ at given $\varepsilon$ and takes a real value for real $z$.

\par
\par
\subsection{Behaviors of the conventional and modified saddle points}
\begin{figure}[t]
\includegraphics[width=0.5\linewidth]{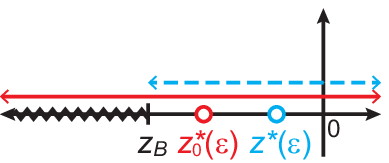}
\caption{(Color online) Diagram of the branch point $z_B$, the branch cut (wiggled line), the conventional saddle point $z_0^* (\varepsilon)$, and the modified saddle point $z^*(\varepsilon)$. The two-headed arrow lines denote the range of $z_0^*$ (thick line) and $z^*$ (dashed line), respectively, as $\varepsilon$ varies.}\label{integration_schematic}
\end{figure}

The conventional saddle point, $z_0^*$, is assumed to be located on the real axis, satisfying
\begin{equation}
H' (z_0^*;\varepsilon)=0
\end{equation}
with $H''(z_0^*;\varepsilon) > 0$.
As $z_0^*$ is a function of $\varepsilon$, thus its location can be varied and pass through the branch point $z_B$
as $\varepsilon$ varies (Fig.~\ref{integration_schematic}).
On the other hand, the modified saddle point $z^*$ satisfies (equation~(\ref{modified_S}))
\begin{equation}
H'(z^*;\varepsilon) - \frac{\alpha}{\tau(z^* -z_B)}=0.\label{appendix_saddle_eq}
\end{equation}
As discussed in Sec.~\ref{sec2} and Fig.~\ref{modified_saddle_point}, there are always two solutions satisfying the above equation
and we choose the saddle point which are located on the right side of $z_B$. Then, it is easy to show
that $z^*>z^*_0$ as well as $z^*>z_B$ from equation~(\ref{two_solutions}) and Fig.~\ref{modified_saddle_point}.
Moreover, for large $\tau$, $z^*$ asymptotically approaches $z_0^*$ when $z_0^* (\varepsilon) > z_B$, while it approaches $z_B$ otherwise.

For convenience, we define $\Delta z=z_0^*(\varepsilon)-z_B$ as an alternative variable for $\varepsilon$ and also
$\delta z_0=z^*-z_0^*$ and $\delta z_B=z^*-z_B$ as asymptotically vanishing positive quantities for $\Delta z>0$ and
$\Delta z\le 0$, respectively. We investigate how these small quantities vanish as a function of $\tau$, in order
to calculate the finite-time correction of the LDF, $\tau^{-1} \ln I(\varepsilon)$, up to the order of $\tau^{-1}$.

First, consider the case $\Delta z>0$. Expansion of equation~(\ref{appendix_saddle_eq}) near $z_0^*$ with small $\delta z_0$
yields
\begin{equation}
H''(z_0^*)~\delta z_0-\frac{\alpha}{\tau (\delta z_0+\Delta z)} = 0, \label{app_expansion1}
\end{equation}
where the $\varepsilon$ dependence is dropped for simplicity and afterwards. The positive solution of the above quadratic equation
is
\begin{equation}
\delta z_0 = -\frac{\Delta z}{2} + \sqrt{\frac{(\Delta z)^2}{4} +\frac{\alpha}{\tau  H''(z_0^*)}  }~, \label{app_delta_z_0}
\end{equation}
which leads to two different scaling behaviors as
\begin{equation}
\delta z_0 = \left\{
\begin{array}{ll}
  \frac{\alpha}{ (\Delta z)H''(z_0^*)}~\tau^{-1}, \quad &\quad \Delta z\sqrt{\tau} \gg 1 \\
  \sqrt{\frac{\alpha}{ H''(z_0^*)}}~\tau^{-1/2}, \quad &\quad 0< \Delta z\sqrt{\tau} \ll 1~.
\end{array} \right. \label{app_delta_z_2}
\end{equation}

Second, consider the case $\Delta z \leq 0$. Expansion of equation~(\ref{appendix_saddle_eq}) near $z_B^*$
with small $\delta z_B$ yields
\begin{equation}
H'(z_B)+H''(z_B)~\delta z_B-\frac{\alpha}{\tau \delta z_B} = 0. \label{app_expansion2}
\end{equation}
Its positive solution is
\begin{equation}
\delta z_B = -\frac{|{\tilde\Delta} z|}{2} + \sqrt{\frac{(\tilde\Delta z)^2}{4} +\frac{\alpha}{\tau  H''(z_B^*)}  }~, \label{app_delta_z}
\end{equation}
where ${\tilde\Delta}z \equiv -H'(z_B)/H''(z_B)$$(\le 0)$ is approximately equal to $\Delta z$ for small $|\Delta z|$.
Again, two scaling regimes are found as
\begin{equation}
\delta z_B = \left\{
\begin{array}{ll}
  \frac{\alpha}{ H'(z_B)}~\tau^{-1}, \quad &\quad |\tilde\Delta z|\sqrt{\tau} \gg 1 \\
  \sqrt{\frac{\alpha}{ H''(z_B)}}~\tau^{-1/2}, \quad &\quad 0< |\tilde\Delta z|\sqrt{\tau} \ll 1~.
\end{array} \right. \label{app_delta_z_3}
\end{equation}

Summarizing equations~(\ref{app_delta_z_2}) and (\ref{app_delta_z_3}), small quantities $\delta z_0$ and $\delta z_B$ are
categorized into three regimes as follows:
\begin{equation}
\left\{
\begin{array}{ll}
 \delta z_0 =\frac{\alpha}{ (\Delta z)H''(z_0^*)}~\tau^{-1}, \quad &\quad \Delta z\sqrt{\tau} \gg 1 \\
 \delta z_0\simeq \delta z_B=  \sqrt{\frac{\alpha}{ H''(z_B)}}~\tau^{-1/2}, \quad &\quad |\Delta z|\sqrt{\tau} \ll 1 \\
  \delta z_B = \frac{\alpha}{ H'(z_B)}~\tau^{-1}, \quad &\quad \Delta z\sqrt{\tau} \ll -1 ~.
 \end{array} \right. \label{app_delta_z_tot}
\end{equation}

\par
\par

\subsection{Integral form of $I(\varepsilon)$}

\begin{figure}[t]
\includegraphics[width=0.7\linewidth]{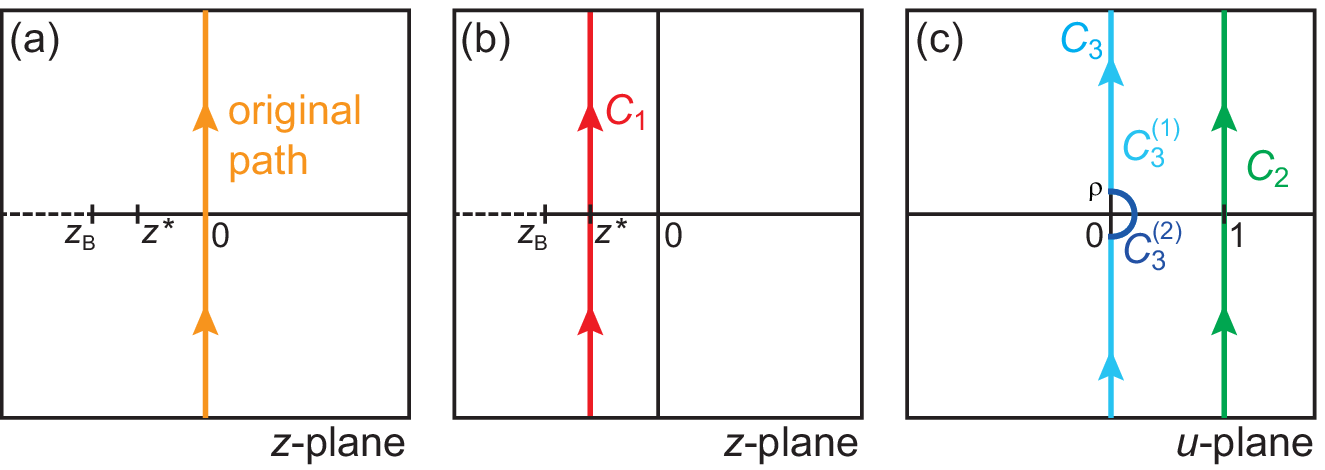}
\caption{(Color online) Diagrams denoting the change of the contour path. (a) Original contour path.
(b) Contour path $C_1$ for equation (\ref{app_int_expansion_modified}).
(c) Contour paths $C_2$ and $C_3$.
}\label{contours_change}
\end{figure}

We calculate equation~(\ref{app_int}) by the steepest descent method along a contour $C_1$
passing through the modified saddle point $z^*$ (Fig.~\ref{contours_change}). By expanding $H(z)$ near $z^*$ up to the second order of
$z-z^*$, we get
\begin{equation}
I(\varepsilon) =g(z^*)~e^{\tau H(z^*)}\int_{C_1} dz~  \frac{e^{\tau H'(z^*)(z-z^*)
+\frac{\tau}{2}H''(z^*)(z-z^*)^2}}{(z-z^*+\delta z_B)^\alpha},\label{app_int_expansion_modified}
\end{equation}
where $H'(z^*)\neq 0$ for the modified saddle-point. The denominator of the integrand was not expanded on purpose because
its expansion series does not converge when $\delta z_B$ is small enough as in equation (\ref{app_delta_z_tot}) for
$\Delta z\sqrt{\tau}\ll 1$.

It is convenient to use a variable $u={1+(z-z^*)/\delta z_B}$. Then, the above integral is rewritten as
\begin{eqnarray}
I(\varepsilon) &=&  g(z^*) (\delta z_B)^{1-\alpha}  \exp[\tau H(z^*)] \int_{C_2} du~ {u^{-\alpha}}\nonumber \\
&\times& \exp \left[ \tau H'(z^*) \delta z_B (u-1) + \frac{\tau}{2}  H''(z^*)(\delta z_B)^2 (u-1)^2 \right]~, \label{arrange1}
\end{eqnarray}
where the contour $C_2$ is given in figure~\ref{contours_change}(c) of the complex $u$ plane.
This integral can be more simplified by replacing the contour $C_2$ by $C_3$ in figure~\ref{contours_change},
which comprises of the vertical sector $C_3^{(1)}$ with $u=iy$ for $|y|>\rho$ and the small half-circular sector
$C_3^{(2)}$ with fixed $\rho$ and $|\phi|<\frac{\pi}{2}$. The half-circular sector $C_3^{(2)}$ is introduced to
avoid the singularity of the integrand at $u=0$.
Later, we will take the $\rho\rightarrow 0$ limit, for simplicity.

Now, consider the case $\Delta z \leq 0$ first, where $\delta z_B$ is small. By expanding $H(z^*)$ and $H'(z^*)$ near $z_B$
up to the order of $(\delta z_B)^2$,
the integral is simplified as
\begin{eqnarray}
I(\varepsilon) &=& g(z_B) (\delta z_B)^{1-\alpha} \exp[\tau H(z_B)] \nonumber \\
 &\times &\int_{C_3} du~ {u^{-\alpha}}\exp \left[ \tau H'(z_B) \delta z_B u + \frac{\tau}{2}
 H''(z_B)(\delta z_B)^2 u^2 \right]~. \label{app_int_C2}
\end{eqnarray}
This is a non-Gaussian integral, which may not be expressed in a closed form in general.

The contribution to the integral from the vertical sector $C_3^{(1)}$ with $u=iy$ for $|y|>\rho$
can be written as
\begin{eqnarray}
I^{(1)}(\varepsilon) &=& 2i g(z_B) (\delta z_B)^{1-\alpha} \exp[\tau H(z_B)] \nonumber \\
&\times& \int_{\rho}^{\infty} dy~y^{-\alpha}  \cos\left[\theta_\alpha+\Theta_B(y)\right] e^{-\frac{\tau}{2}
H''(z_B) (\delta z_B)^2 y^2} , \label{app_int_C3}
\end{eqnarray}
where $\theta_\alpha= \frac{\pi}{2}\alpha$ and $\Theta_B(y) = -\tau H'(z_B) \delta z_B y$.
The second contribution to the integral from the half-circular sector $C_3^{(2)}$ with small $\rho$ and $|\phi|<\frac{\pi}{2}$
is given as
\begin{eqnarray}
I^{(2)}(\varepsilon) &=& 2i g(z_B) (\delta z_B)^{1-\alpha} \exp[\tau H(z_B)] \nonumber \\
&\times& \rho^{1-\alpha} \left[\frac{\cos\theta_\alpha}{1-\alpha}
- \Theta_B(\rho) \frac{\sin\theta_\alpha}{2-\alpha} +\cdots\right], \label{app_int_C3_2}
\end{eqnarray}
up to the order of $\rho^{2-\alpha}$. The integral $I(\varepsilon)=I^{(1)}(\varepsilon)
+I^{(2)}(\varepsilon) $ should be independent of $\rho$.

Note that  $I^{(2)}(\varepsilon)$ vanishes in the $\rho\rightarrow 0$ limit  for $\alpha < 1$, so
$I(\varepsilon)=\lim_{\rho\rightarrow 0} I^{(1)}(\varepsilon)$.
For general $\alpha \ge 1$, both $I^{(1)}(\varepsilon)$ and $I^{(2)}(\varepsilon)$ diverge, but
the divergent terms cancel out exactly, leading to a convergent value for the integral $I(\varepsilon)$.
This cancelation can be seen easily by integrating equation (\ref{app_int_C3}) by parts.
For $1<\alpha<2$, the sum of $I^{(1)}$ and $I^{(2)}$ yields that
\begin{eqnarray}
I(\varepsilon) &=& 2i g(z_B) (\delta z_B)^{1-\alpha} \exp[\tau H(z_B)] \nonumber \\
&\times& \int_{0}^{\infty} dy~\frac{y^{1-\alpha}}{\alpha-1}  \frac{d}{dy}
\left[\cos\left[\theta_\alpha+\Theta_B(y)\right] e^{-\frac{\tau}{2}
H''(z_B) (\delta z_B)^2 y^2}\right] . \label{app_int_C3_3}
\end{eqnarray}
At $\alpha=1$, ${y^{1-\alpha}}/{(\alpha-1)}$ in the integrand should be replaced by $-\ln y$.
For a larger value of $\alpha$, we only need to perform the appropriate number of successive integrations
of equation (\ref{app_int_C3})
by parts.

Now, consider the case $\Delta z > 0$ where $\delta z_0$ is small. By expanding $H(z^*)$ and $H'(z^*)$ near $z_0$
up to the order of $(\delta z_0)^2$, the integral becomes
\begin{eqnarray}
I(\varepsilon) &=& g(z_0^*) (\delta z_B)^{1-\alpha}  \exp\left[\tau H(z_0^*)+\frac{\tau}{2} H''(z_0^*)(\Delta z)^2 \right]
\int_{C_3} du~ u^{-\alpha}\nonumber \\
&\times& \exp \left[ -\tau H''(z_0^*) \delta z_B (\Delta z) u + \frac{\tau}{2}  H''(z_0^*)(\delta z_B)^2 u^2 \right].
\label{app_int_case2}
\end{eqnarray}
Then, the two contributions to the integral are written as
\begin{eqnarray}
I^{(1)}(\varepsilon) &=& 2i g(z_0^*) (\delta z_B)^{1-\alpha} \exp\left[\tau H(z_0^*)+\frac{\tau}{2} H''(z_0^*)(\Delta z)^2\right] \nonumber \\
&\times& \int_{\rho}^{\infty} dy~y^{-\alpha}  \cos\left[\theta_\alpha +\Theta_0(y)\right] e^{-\frac{\tau}{2}
H''(z_0^*) (\delta z_B)^2 y^2}, \label{app_int_C32}\\
I^{(2)}(\varepsilon) &=& 2i g(z_0^*) (\delta z_B)^{1-\alpha}  \exp\left[\tau H(z_0^*)+\frac{\tau}{2} H''(z_0^*)(\Delta z)^2\right]\nonumber \\
&\times& \rho^{1-\alpha} \left[\frac{\cos\theta_\alpha}{1-\alpha}
- \Theta_0(\rho) \frac{\sin\theta_\alpha}{2-\alpha} +\cdots\right],
\label{app_int_C33}
\end{eqnarray}
where $\Theta_0(y) = \tau H''(z_0^*) (\Delta z)\delta z_B  y$.
Again, for $\alpha < 1$, $I(\varepsilon)=\lim_{\rho\rightarrow 0} I^{(1)}(\varepsilon)$, while
a similar calculation as above can be done for $\alpha\ge 1$.

\begin{figure}[t]
\includegraphics[width=0.9\linewidth]{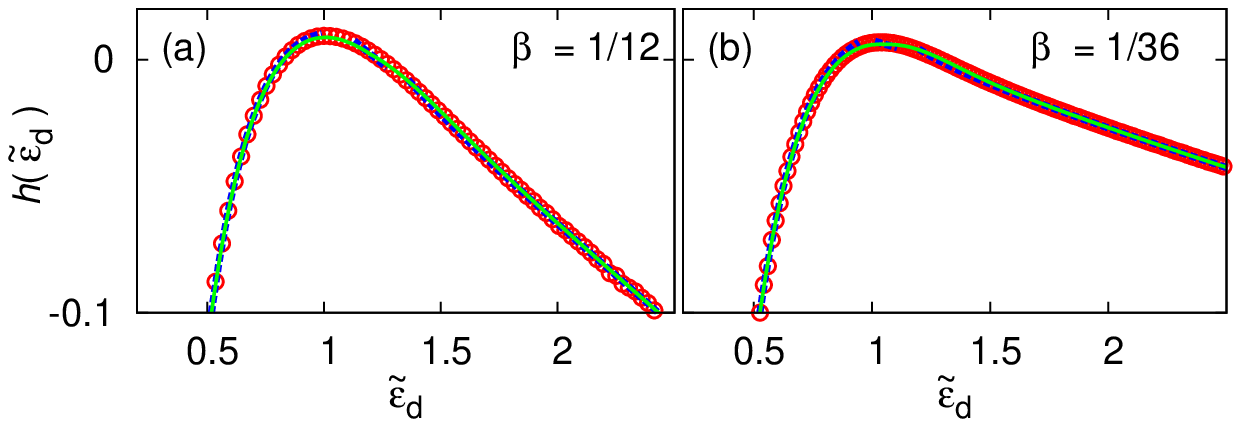}
\caption{(Color online) LDF curves for the dissipated power $\tilde\varepsilon _d$ with finite-time corrections
at $\tau=100$ with (a) $\beta=1/12$ and (b) $\beta=1/36$. Red-open circles, blue-dashed lines, and green solid lines denote
numerical simulation data, piece-wise analytic curves (closed form in \ref{closed}),
and full-range analytic curves (integral form), respectively.
}\label{app_full_range_LDF_curves}
\end{figure}

Note that equations (\ref{app_int_C3})-(\ref{app_int_C3_3}) and (\ref{app_int_case2})-(\ref{app_int_C33})
become identical for small $|\Delta z|$, which implies that the integral $I(\varepsilon)$ also has only three different scaling regimes,
similar to the small quantities $\delta z_0$ and $\delta z_B$ in equation (\ref{app_delta_z_tot}).
Our results together with equations~(\ref{app_delta_z_0}) and (\ref{app_delta_z})
for $\delta z_0$ and $\delta z_B$ provide us to calculate $I(\varepsilon)$ or the LDF's numerically with a very high precision
even in the rare-event regime (large $|\varepsilon|$). As an example, we numerically calculated the integrals of
the dissipated power for $\beta<1/2$ studied in the main text, where $\alpha=1/2$, $H(z;\varepsilon)=(\varepsilon z +1 -\sqrt{1+2z})/2$
 ($\gamma=1$), $z_B=-2\beta(1-\beta)$, and $g(z)$ is given by equation (\ref{g_lambda1}).
 As shown in Fig.~\ref{app_full_range_LDF_curves}, the numerical integrations show the perfect match with
 the numerical simulation data at $\tau=100$, obtained from direct integration of the Langevin equation (\ref{Langevin}).

\subsection{Closed form of $I(\varepsilon)$ in the three scaling regimes}\label{closed}

We assume that $\alpha<1$ in this subsection, for simplicity. Generalization to general $\alpha$ is straightforward.
First, consider the case $ \Delta z\sqrt{\tau} \gg 1$. From equation~(\ref{app_delta_z_tot}), we find $\delta z_0 \sim \tau^{-1}$,
so $\delta z_B=\delta z_0+\Delta z\approx \Delta z$.
In equation (\ref{app_int_C32}), the integral can be done easily by noting that
$\int_0^\infty ds~s^{-\alpha} \exp [-\tau (s\pm i)^2]\approx \frac{1}{2}  (\pm i)^\alpha\sqrt{\pi/\tau}$ for large $\tau$,
which leads to
\begin{equation}
I(\varepsilon)=i\left[\frac{2\pi}{\tau  H'' (z_0^*)}\right]^{1/2}  \frac{g(z_0^*)}{ (\Delta z)^\alpha}
e^{\tau H(z_0^*)}.\label{app_conventional_finite_int}
\end{equation}
As expected, we recover the result by using the conventional saddle point method in equation (\ref{conventionalSP}),
which should be correct for any positive $\alpha$.

Second, consider the case $ |\Delta z|\sqrt{\tau} \ll 1$, where $\delta z_0 \simeq \delta z_B \sim \tau^{-1/2}$
from equation~(\ref{app_delta_z_tot}), so
$\Delta z$ can be ignored. In this case,  $\Theta_B(y)\approx\Theta_0(y) \approx \tau H''(z_0^*) (\Delta z) \delta z_B y$ becomes
negligible for large $\tau$. In addition, $\tau (\delta z_B)^2 \approx  \alpha/ H''(z_B)$ from equation~(\ref{app_delta_z_tot}).
Then, equations~(\ref{app_int_C3}) and (\ref{app_int_C32}) are simplified as
the following identical formula in the $\rho\rightarrow 0$ limit;
\begin{eqnarray}
I(\varepsilon) &=& 2i g(z_B) (\delta z_B)^{1-\alpha} e^{\tau H(z_B)} \cos \theta_\alpha
 \int_{0}^{\infty} dy ~ y^{-\alpha}  e^{-\frac{\alpha}{2}y^2} \nonumber \\
&=&i \frac{2^{\frac{1-\alpha}{2}}\Gamma(\frac{1-\alpha}{2}) \cos (\frac{\pi\alpha}{2})  g(z_B) }
{\left[\tau H''(z_B)\right]^{\frac{1-\alpha}{2}}} ~e^{\tau H(z_B)}. \label{app_near_zB_result1}
\end{eqnarray}
For the last equality, $\int_0^\infty dy ~y^{-\alpha} e^{-\alpha y^2/2}=2^{-(1+\alpha)/2}\alpha^{-(1-\alpha)/2}\Gamma(\frac{1-\alpha}{2})$ is used.

Finally, consider the case $  \Delta z\sqrt{\tau} \ll -1$, where $\delta z_B = \frac{\alpha}{H'(z_B)} \tau^{-1}$
from equation~(\ref{app_delta_z_tot}). In this case, $\Theta_B(y) \approx -\alpha y$ and
$\tau H''(z_B) (\delta z_B)^2 \sim \tau^{-1}$ is negligible.
Then, equation~(\ref{app_int_C3}) in the $\rho\rightarrow 0$ limit is simplified as
\begin{eqnarray}
I(\varepsilon)&=& \frac{2i g(z_B)\alpha^{1-\alpha}}{\left[\tau H'(z_B)\right]^{1-\alpha}}~  e^{\tau H(z_B)}
\int_0^\infty dy ~y^{-\alpha}
\cos \left(\theta_\alpha -\alpha y\right)   \nonumber \\
&=& i \frac{2 \Gamma(1-\alpha) \sin (\alpha \pi) g(z_B)}{\left[\tau H'(z_B)\right]^{1-\alpha}}  e^{\tau H(z_B)}
\label{app_near_zB_result2}
\end{eqnarray}
For the last equality, $\int_0^\infty dy ~y^{-\alpha}~ e^{i(\alpha y)}=\alpha^{-(1-\alpha)} \Gamma(1-\alpha)~
e^{i\frac{\pi}{2}(1-\alpha)}$ is used.

Summarizing equations.~(\ref{app_conventional_finite_int}), (\ref{app_near_zB_result1}), and (\ref{app_near_zB_result2})
for $\alpha=1/2$, $I(\varepsilon)$ becomes
\begin{equation}
I(\varepsilon) = \left\{
\begin{array}{ll}
  i \sqrt{\frac{2\pi}{\tau(\Delta z)  H'' (z_0^*)}}  ~g(z_0^*)
~e^{\tau H(z_0^*)}, & \Delta z \sqrt{\tau} \gg 1\\
  i\frac{\Gamma(1/4) }{\left[2\tau H''(z_B)\right]^{1/4}} ~g(z_B) ~e^{\tau H(z_B)}, &  |\Delta z| \sqrt{\tau} \ll 1\\
  i \frac{2 \sqrt{\pi} }{\sqrt{\tau H'(z_B)}} ~g(z_B) ~e^{\tau H(z_B)}, &  \Delta z \sqrt{\tau} \ll -1 \\
\end{array} \right. \label{app_I_tot1}
\end{equation}
where $\Delta z=z^*_0-z_B$.
The above equation corresponds to the result in reference~\cite{Sabhapandit1}. When the branch cut is located to its
right side of the branch point $z_B$ with a singularity like $(z_B-z)^{-\alpha}$ in equation (\ref{app_int}), we get
the same equation for $I(\varepsilon)$ except for changing $H'$ to $-H'$ and $\Delta z$ to $-\Delta z$.


\begin{thebibliography}{10}
\bibitem{note1} The generating function or equation~(\ref{Fourier_transform}) is properly defined only in a region of convergence on the complex $\lambda$ plane, so all singular points of the generating function must be located at the boundary or the outside of the convergence region. Thus, the phrase that we are using in this paper such as `the generating function has a singularity' makes sense only when the analytic continuation of the generating function is taken into account.
\end{thebibliography}
\end{document}